\documentclass[fleqn,usenatbib]{mnras}
\usepackage[T1]{fontenc}
\usepackage{graphicx} 
\usepackage{amsmath}
\usepackage{amssymb}
\usepackage{color}
\usepackage{multicol}
\usepackage{threeparttable}
\usepackage{newtxtext,newtxmath}

\DeclareRobustCommand{\VAN}[3]{#2}
\let\VANthebibliography\thebibliography
\def\thebibliography{\DeclareRobustCommand{\VAN}[3]{##3}\VANthebibliography}

\def\refjnl#1{{\rm#1}}
\def\aj{\refjnl{AJ}}                   
\def\apj{\refjnl{ApJ}}                 
\def\apjl{\refjnl{ApJ}}                
\def\apjs{\refjnl{ApJS}}               
\def\ao{\refjnl{Appl.~Opt.}}           
\def\aap{\refjnl{A\&A}}                
\def\mnras{\refjnl{MNRAS}}             
\def\nar{\refjnl{New A Rev.}}          
\def\pasa{\refjnl{PASA}}               
\def\pasp{\refjnl{PASP}}               
\def\pasj{\refjnl{PASJ}}               
\def\nat{\refjnl{Nature}}              
\def\physrep{\refjnl{Phys.~Rep.}}   

\def\be{\begin{equation}} 
\def\ee{\end{equation}}

\def\msun{{\Msun}}

\def\HI{\hbox{H~$\scriptstyle\rm I\ $}}

\def\gsim{\lower.5ex\hbox{\gtsima}} 
\def\lsim{\lower.5ex\hbox{\ltsima}} \def\gtsima{$\; \buildrel > \over 
\sim \;$} \def\ltsima{$\; \buildrel < \over \sim \;$} \def\prosima{$\; 
\buildrel \propto \over \sim \;$} \def\gsim{\lower.5ex\hbox{\gtsima}} 
\def\lsim{\lower.5ex\hbox{\ltsima}} 
\def\simgt{\lower.5ex\hbox{\gtsima}} 
\def\simlt{\lower.5ex\hbox{\ltsima}} 
\def\simpr{\lower.5ex\hbox{\prosima}} \def\la{\lsim} \def\ga{\gsim} 
  
 \def\gtsima{$\; \buildrel > \over \sim \;$} 
\def\ltsima{$\; \buildrel < \over \sim \;$} 
\def\gsim{\lower.5ex\hbox{\gtsima}} 
\def\lsim{\lower.5ex\hbox{\ltsima}} 
\def\simgt{\lower.5ex\hbox{\gtsima}} 
\def\simlt{\lower.5ex\hbox{\ltsima}} 
\def\simpr{\lower.5ex\hbox{\prosima}} 
\def\la{\lsim} 
\def\ga{\gsim}

\def\msun{\,{\rm \Msun}}

\def\E3{{\cal E}_{\rm g}^{III}}

\def\Msun{\rm M_\odot}

\def\Zsun{\rm Z_\odot}

\def\Msun{\rm M_\odot}

\def\Zsun{\rm Z_\odot}
\def\M*{M_*}

\def\Z*{Z_*}
\def\L*{L_*}

\def\feff{f_*^{eff}}

\def\feff{f_\star^\mathrm{eff}}
\def\fesc{f_\mathrm{esc}}

\title[Galaxy formation in the EoR]{Astraeus I: The interplay between galaxy formation and reionization} 
\author[Hutter et al.]{Anne Hutter$^{1}$\thanks{a.k.hutter@rug.nl}, Pratika Dayal$^1$, Gustavo Yepes $^{2,3}$, Stefan Gottl\"ober $^4$, Laurent Legrand$^1$, Graziano Ucci$^1$\\
$^{{1}}$ Kapteyn Astronomical Institute, University of Groningen, P.O. Box 800, 9700 AV Groningen, The Netherlands \\
$^{2}$ Departamento de Fısica Teorica, Modulo 8, Facultad de Ciencias, Universidad Autonoma de Madrid, 28049 Madrid, Spain\\
$^{3}$ CIAFF, Facultad de Ciencias, Universidad Autonoma de Madrid, 28049 Madrid, Spain\\
$^{4}$ Leibniz-Institut f\"ur Astrophysik, An der Sternwarte 16, 14482 Potsdam, Germany
}

\date{Accepted 2021 February 25. Received 2021 February 25; in original form 2020 April 17}

\pubyear{2021}

\begin{document}
\label{firstpage}
\pagerange{\pageref{firstpage}--\pageref{lastpage}}
\maketitle

\begin{abstract}
We introduce a new self-consistent model of galaxy evolution and reionization, {\sc astraeus} (semi-numerical r{\bf A}diative tran{\bf S}fer coupling of galaxy forma{\bf T}ion and {\bf R}eionization in N-body d{\bf A}rk matt{\bf E}r sim{\bf U}lation{\bf S}), which couples a state-of-the-art N-body simulation with the semi-analytical galaxy evolution {\sc delphi} and the semi-numerical reionization scheme {\sc cifog}. {\sc astraeus} includes all the key processes of galaxy formation and evolution (including accretion, mergers, supernova and radiative feedback) and follows the time and spatial evolution of the ionized regions in the intergalactic medium (IGM). Importantly, it explores different radiative feedback models that cover the physically plausible parameter space, ranging from a weak and delayed to a strong and immediate reduction of gas mass available for star formation. From our simulation suite that covers the different radiative feedback prescriptions and ionization topologies, we find that radiative feedback continuously reduces star formation in galaxies with $M_h\lesssim10^{9.5}\msun$ upon local reionization; larger mass halos are unaffected even for the strongest and immediate radiative feedback cases during reionization. For this reason, the ionization topologies of different radiative feedback scenarios differ only on scales smaller than $1-2$~comoving Mpc, and significant deviations are only found when physical parameters (e.g. the escape fraction of ionizing photons) are altered based on galactic properties. Finally, we find observables (the ultra-violet luminosity function, stellar mass function, reionization histories and ionization topologies) are hardly affected by the choice of the used stellar population synthesis models that either model single stars or binaries.
\end{abstract}


\section{Introduction}
\label{sec_introduction}

The Epoch of Reionization (EoR) represents the last major phase transition of hydrogen in the history of the Universe. Its beginning is marked by the appearance of the first stars and galaxies, whose Lyman continuum photons (with energy $E>13.6$eV) gradually ionize the neutral hydrogen (\HI) in the intergalactic medium (IGM). The growing ionized bubbles around galaxies merge and expand until the IGM is completely ionized by $z\simeq6$ \citep[e.g.][]{Fan2006, Becker2015}. A rising number of high-redshift galaxy observations are providing us with increasing hints on the properties and numbers of star-formation driven ionizing sources \citep[e.g.][]{smit2014, Bouwens2015, Smit2018, Ouchi2018, DeBarros2019, Maseda2020}. 
These galaxy data-sets are complemented by (upper limits on) the 21cm emission from \HI in the IGM during reionization obtained by experiments such as LOFAR\footnote{Low Frequency Array,  \url{http://www.lofar.org}} \citep{Patil2017, Mertens2020}, MWA\footnote{Murchison Widefield Array, \url{http://www.mwatelescope.org}} \citep{Li2019, Barry2019} and PAPER\footnote{Precision Array for Probing EoR, \url{http://eor.berkeley.edu}}: \citep{Kolopanis2019}.
Over the next decade, this 21cm data will be supplemented by that from state-of-the-art radio interferometers, such as the Square Kilometre Array \citep[SKA;][]{Carilli2004} and the Hydrogen Epoch of Reionization Array \citep[HERA;][]{DeBoer2017}, which are designed to measure the temporal and spatial evolution of the ionized regions, i.e. the reionization topology \citep[e.g.][]{Greig2019, Seiler2019, Elbers2019, Hutter2017, Hutter2020}. Despite this progress, the reionization topology, the properties of the ionizing sources and the impact of reionization on the evolution of galaxy properties through radiative feedback effects remain key outstanding questions in the field of physical cosmology \citep[for a review see e.g.][]{dayal2018}.

As the IGM becomes ionized, the associated ultra-violet background (UVB) photo-heats the gas in halos and the IGM to about $\sim10^4$K. The higher temperature and rising pressure of the gas in a halo causes a fraction of the gas to photo-evaporate into the IGM \citep{barkana-loeb1999, shapiro2004} and raises the Jeans mass for galaxy formation \citep[reducing the amount of gas being accreted;][]{couchman-rees1986, efstathiou1992, Hoeft2006}. Both mechanisms lead to a reduction of gas mass and the associated star formation rate, particularly in low-mass halos.
However, modelling the impact of reionization feedback on galaxy formation remains challenging due to its complex dependence on halo mass and redshift, the patchiness and strength of the UVB and the redshift at which an assembling halo is first irradiated by the UVB \citep[e.g.][]{Gnedin2000, sobacchi2013b}. 

Early works have studied the effects of radiative (photoheating) feedback on galaxies in cosmological hydrodynamical simulations by quantifying the loss of baryons in low-mass halos in the presence of a homogeneous UVB \citep[e.g.][]{Hoeft2006, Okamoto2008, Naoz2013}. However, since reionization is a spatially inhomogeneous and temporal extended process, an increasing number of radiation hydrodynamical simulations have studied the impact of an inhomogeneous and evolving UVB on the galaxy population and found a reduction in the star formation rates in low-mass galaxies with halo mass $M_h\lesssim10^9\msun$ \citep{Gnedin2000, Hasegawa2013, Gnedin2014, Pawlik2015, ocvirk2016, Ocvirk2018, Katz2019, Wu2019}. Most importantly, a number of such radiation hydrodynamical simulations show that the star-formation-suppressing effect of radiative increases with time, even after the Universe has been mostly ionized \citep{Gnedin2014, ocvirk2016, Ocvirk2018, Wu2019}, which could be attributed to a decrease in self-shielding and a slower heating of the gas \citep{Wu2019}. The suppression of star formation is also found to be dependent on the environment, i.e. galaxies in over-dense regions that ionize earlier feature higher star formation rates which declines sharply after local reionization for low-mass halos with $M_h<10^9\msun$ \citep{Dawoodbhoy2018}. Highlighting the interplay between galaxy formation and reionization, \citet{Wu2019} have shown that a stronger stellar feedback reduces the star formation within the galaxy and hence the UVB, weakening the strength of radiative feedback.
In order to investigate the signatures of radiative feedback on the ionization topology, a number of works have combined N-body simulations with radiative transfer and used different suppression models for the ionizing emissivities of low-mass halos \citep[e.g.][]{iliev2007, iliev2012, Dixon2016}. However, since these simulations do not contain a galaxy evolution model, the gas mass in halos below the local Jeans mass of the photo-heated IGM is instantaneously suppressed in ionized regions. Different suppression models mostly affect the timing of reionization as compared to the ionization topology \citep{Dixon2016}.

In this paper, our aim is to quantify the effects of radiative feedback, both, on the underlying galaxy population as well the ionization topology during the EoR to answer questions including: When and which galaxies are most affected by radiative feedback? Is the patchiness of reionization imprinted in galaxy observables? How does radiative feedback impact high-redshift observables (including the UV luminosity function, stellar mass function and the redshift evolution of the star formation rate density and stellar mass density) and the 21cm signal from the neutral regions in the IGM? This naturally requires coupling galaxy formation and reionization using large volume simulations with a high-resolution to be able to study the ionization histories of galaxies based, both, on their masses as well as their location in the cosmic web. For example, in an inside-out reionization scenario, low-mass galaxies, can either be located in high-density regions that get ionized quite early on (therefore being strongly affected by UVB feedback) or in low-density regions that are ionized later (resulting in weak to no UVB feedback). 

For this reason, we have built the {\sc astraeus} (semi-numerical r{\bf A}diative tran{\bf S}fer coupling of galaxy forma{\bf T}ion and {\bf R}eionization in N-body d{\bf A}rk matt{\bf E}r sim{\bf U}lation{\bf S}) framework that self-consistently couples a state-of-the-art N-body simulation (very small multi-dark; {\sc vsmd}) with a semi-analytic model of galaxy formation \citep[{\sc delphi};][] {dayal2014,dayal2015,dayal2017a} and a semi-numerical reionization scheme \citep[{\sc cifog};][]{Hutter2018}. While similar approaches have been followed for {\sc meraxes} \citep{Mutch2016} and {\sc rsage} \citep{Seiler2019}, these works have only focused on exploring the suppression of gas mass and star formation in low-mass halos based on 1D radiation hydrodynamical simulations \citep{sobacchi2013b}. In contrast, in this paper we explore different radiative feedback scenarios that range from a minimum one with gas loss via the characteristic mass approach outlined in \citet{Gnedin2000} to a maximum one where the amount of gas is instantaneously reduced in halos with masses below the local Jeans mass of the ionized region; although similar in spirit to the work of \citet{iliev2012} and \citet{Dixon2016}, our model is an advancement on these works given it uses a much more sophisticated model for galaxy formation and the associated ionizing emissivity.
Besides its key strength of supporting multiple radiative feedback models, {\sc astraeus} comprises (1) a large volume and high-resolution N-body simulation that allows us to simultaneously explore the large-scale reionization topology whilst resolving sources down to the atomic cooling mass at $z\sim6$, (2) a galaxy formation model that uses only three free parameters with feedback being linked to the underlying halo potential, and (3) supports multiple models for the ionizing escape fraction $f_\mathrm{esc}$ that enable us to cover the physically plausible range of reionization scenarios.

The paper is structured as follows. In Section \ref{sec_model} we describe the underlying N-body simulation and the theoretical galaxy and reionization model, as well as our different models of radiative feedback. In Section \ref{sec_fitting} we compare our results to observational constraints, such as the luminosity and stellar mass functions, and the Thomson optical depth for reionization. We then use our different models for radiative feedback to investigate how the strength and timing of the suppression of star formation in a galaxy depends on its gravitational potential and local reionization in Section \ref{sec_SFR}, how radiative feedback affects the ionization topology and thus the power spectrum of the 21cm signal in Section \ref{sec_21cm_ps}, and whether assuming a different stellar population synthesis models affects any observables in Section \ref{sec_sps}. We conclude in Section \ref{sec_conclusions}.

\section{The theoretical model}
\label{sec_model}

In this Section, we describe our self-consistent, semi-numerical model that couples high-redshift galaxy formation and reionization, {\sc astraeus}\footnote{{\sc astraeus} can be built from the source code publicly available under \url{https://github.com/annehutter/astraeus}. {\sc astraeus} includes a new implementation of {\sc delphi} and uses the {\sc cifog} library.}. Using the evolving DM density distribution from a high-resolution N-body simulation (Section \ref{subsec_nbody}), {\sc astraeus} couples an enhanced version of the semi-analytic galaxy evolution model {\sc delphi} \citep[][Section \ref{subsec_gal_model}]{dayal2014} to the semi-numerical reionization code {\sc cifog}\footnote{{\sc cifog} is publicly available under \url{https://github.com/annehutter/grid-model}.} \citep[][Section \ref{subsec_reion_model}]{Hutter2018}. The key novelty of {\sc astraeus} is that it allows us to explore a wide range of scenarios for the interplay between galaxy formation and reionization using a minimum number of mass- and redshift-independent free parameters.

\subsection{N-body simulation}
\label{subsec_nbody}

In this work we use the high resolution {\it Very Small MultiDark Planck} {\sc (vsmdpl)} N-body simulation, performed as part of the {\sc multidark} simulation project\footnote{See {\tt www.cosmosim.org} for further information about the Multidark suite of simulations  and access to the simulations database.}. This new simulation, with a box size of $160 h^{-1}$ comoving Mpc (cMpc), was run with the same number of particles ($3840^3$) and using the same {\sc gadget-2} Tree+PM N-body code \citep{springel2005} as in the other Multidark simulations described in \citet{Klypin2016}. We also used the same cosmological parameters to set up initial conditions, namely $[\Omega_\Lambda, \Omega_m, \Omega_b, h, n_s, \sigma_8]$ = $[0.69,~ 0.31, ~0.048, ~0.68, ~0.96, ~0.83]$. The Zeldovich approximation was used to produce the particle positions and velocities at an initial redshift of $z=150$. The mass per dark matter particle is $6.2 \times 10^6 h^{-1}\, \msun$ and the equivalent Plummer's gravitational softening was set to $2h^{-1}$ comoving kpc at $z>1$. A total of 150 different snapshots of the simulation, equally spaced in expansion factor, were stored from $z=25$ until $z=0$, with $63$ snapshots covering the redshifts $z=25$ to $z=6$. The \textsc{Rockstar}  phase-space  halo finder \citep{behroozi2013_rs} was used to identify  all halos and subhalos in each of the 150 snapshots, down to a minimum of 20 particles per halo resulting in a minimum resolved halo mass of $1.24 \times 10^8 h^{-1}\, \msun$. In addition, merger trees from the {\sc rockstar} halo catalogues were computed using the {\sc consistent trees} \citep{behroozi2013_trees} method. 
While the vertical merger trees obtained from {\sc consistent trees} are well suited to follow the evolution history of a single galaxy, i.e. following the evolution of the progenitors of a galaxy, they do not track the galaxy population on a redshift-step-by-redshift-step basis as required for reionization. In order to use {\sc astraeus} as a semi-analytic galaxy formation code run on a tree-branch-by-tree-branch basis (i.e. fully vertical) as in {\sc sage} \citep{Croton2016} or {\sc delphi} \citep{dayal2014} or on a redshift-by-redshift basis (i.e. fully horizontal) as in {\sc meraxes} \citep{Mutch2016}, we re-sort the {\sc consistent tree} outputs as follows. We keep the merger-tree-by-merger-tree order but {\it each} merger tree is sorted by redshift (horizontally sorted), and we refer to this as locally-horizontally sorted. This sorting allows us to include the impact of ``horizontal" processes such as reionization for galaxies at a given timestep before they are evolved to the successive redshift snapshot (see Appendix \ref{app_merger_trees} for details). However, we refrain from generating fully horizontal outputs on a galaxy-by-galaxy basis, as such an order would impede the possibility of following the evolution of a single galaxy easily and limit the flexibility of {\sc astraeus} to be used for non-reionization galaxy studies in the future. 

In the following we run {\sc astraeus} on the full merger trees but limit our discussion of galactic properties to halos where these properties have converged. As shown in Appendix \ref{app_resolution}, we find that our model converges for halos with a DM mass of $M_h\geq10^{8.6}\msun$ (corresponding to halos with at least 50 particles).

\subsection{Semi-analytic galaxy modelling}
\label{subsec_gal_model}

Our semi-analytic galaxy formation model includes all the key baryonic processes of gas accretion, gas and stellar mass being brought in by mergers, star formation and the associated supernovae (SN) feedback and radiative feedback from reionization. At each time step, these are coupled to the merger- and accretion-driven growth of the dark matter halos obtained from the N-body simulations as explained in this section. Throughout this work, we use a Salpeter \citep{salpeter1955} initial mass function (IMF) with a slope of $\gamma=2.35$ between $0.1-100~\Msun$.

\subsubsection{Gas accretion and mergers}
\label{subsubsec_gas}

There are two ways in which a galaxy can build up its gas content: through smooth accretion from the IGM and through mergers in the case that a galaxy has progenitors. On the one hand, at the beginning of a time step, a galaxy of halo mass $M_h(z)$ that has no progenitors, can, in principle, smoothly accrete an initial gas mass, $M_g^i(z)$, corresponding to the cosmological baryon-to-dark matter ratio such that $M_g^i(z) = (\Omega_b/\Omega_m) M_h(z)$. However, reionization feedback can reduce the initial gas mass by photo-evaporating gas out of the potential. In this case $M_g^i(z) = f_g (\Omega_b/\Omega_m) M_h(z)$ where $f_g$ is the gas fraction that remains available for star formation in the presence of an UVB as explained in Sec. \ref{subsec_reion_model} that follows. 

On the other hand, galaxies that have (say $N_p$) progenitors can also gain gas through mergers. In this case, the merged gas mass can be expressed as
\begin{eqnarray}
M_\mathrm{g}^\mathrm{mer}(z) & =  & \sum_{p=1}^{N_p} M_{\mathrm{g},p}(z + \Delta z),
\end{eqnarray}
where $M_{g,p}(z + \Delta z)$ is the final gas mass of the previous time step brought in by the merging progenitors of halo mass $M_{h,p}(z + \Delta z)$. The accreted gas mass in this case is given by
\begin{eqnarray}
M_\mathrm{g}^\mathrm{acc}(z) & = & \frac{\Omega_b}{\Omega_m} \left[M_h(z) - \sum_{p=1}^{N_p} M_{h,p}(z + \Delta z) \right],
\end{eqnarray}
where we have made the reasonable assumption that accretion of halo mass from the IGM drags in a cosmological fraction of gas mass. 

Accounting for the impact of reionization feedback, the initial gas mass can be expressed as
\begin{eqnarray}
M_\mathrm{g}^i(z) &=& \min \left[M_\mathrm{g}^\mathrm{mer}(z)+M_\mathrm{g}^\mathrm{acc}(z),\ f_g \frac{\Omega_b}{\Omega_m}M_h(z) \right].
\label{eq_MgasIni}
\end{eqnarray}

\subsubsection{Star formation and stellar mass assembly}
\label{subsubsec_star_formation}

We assume that at a given time step this initial gas mass, $M_\mathrm{g}^i$, can form stars with an {\it effective} efficiency ($\feff$) which is the minimum between that required to eject the rest of the gas from the halo potential ($f_\star^\mathrm{ej}$) and quench star formation and an upper limit ($f_\star \sim 1-3\%$) such that $\feff = \min\left[ f_\star, f_\star^\mathrm{ej} \right]$; details of the calculation of $f_\star^\mathrm{ej}$ follow in Section \ref{subsubsec_SN_feedback}. The newly formed stellar mass at any time step can then be expressed as 
\begin{eqnarray}
 M_\star^\mathrm{new}(z) &=& \feff M_\mathrm{g}^i (z).
 \label{eq_Mstar_new}
\end{eqnarray}
Physically, the effective efficiency can be thought of as $\feff = f_s/t_s$ i.e. a fraction ($f_s$) of the gas mass that can form stars over a timescale $t_s$\footnote{With the time steps in the {\sc vsmdpl} simulation scaling with the logarithm of the scale factor and hence increasing towards lower redshifts, the actual star formation efficiency increases towards higher redshifts. However, we note that this effect is not major as the time steps during the Epoch of Reionization range from $17-30$~Myrs at $z\simeq10-6$ with deviations being around $\sim30\%$ from a constant time step of $23.5$~Myrs (corresponding to the time step at the midpoint of reionization at $z\simeq7$ in {\sc vsmdpl}).}. Given that $\feff$ is linked to the underlying halo potential, our model results in low-mass galaxies ($M_h \lsim 10^{9.3}\msun$ at $z=5$) being {\it star formation efficiency limited} with $\feff = f_\star^\mathrm{ej}$, while larger mass halos form stars with a constant efficiency $\feff = f_\star$ \citep[see also][]{dayal2014}. 

In addition, stellar mass can also be brought in by merging progenitors ($M_{\star,p}$) such that
\begin{eqnarray}
 M_\star^\mathrm{mer}(z) &=& \sum_{p=1}^{N_p} M_{\star,p}(z + \Delta z),
 \label{eq_Mstar_mer}
\end{eqnarray}
resulting in a total stellar mass 
\begin{eqnarray}
 M_\star(z) &=& M_\star^\mathrm{new}(z) + M_\star^\mathrm{mer}(z).
 \label{eq_Mstar}
\end{eqnarray}
We note that our model does not explicitly describe starbursts triggered by mergers \citep[as e.g. in][]{Croton2016} and aim to assess the relative role of mergers and accretion in the gas and stellar mass assembly in a future work.

Star formation in galaxies has two key physical effects: firstly, at the end of their life, high-mass stars explode as Type II supernovae (SNII) which can eject gas mass (say, $M_\mathrm{g}^\mathrm{ej}$) from the galaxy. Secondly, star formation provides \HI ionizing photons; the fraction of these photons that can escape into the IGM ($\fesc$) contribute to reionizing and heating the IGM as detailed in Sec. \ref{subsec_reion_model}.

\subsubsection{Supernova feedback}
\label{subsubsec_SN_feedback}
The explosion of high-mass stars as SNII injects thermal and kinetic energy into the interstellar medium (ISM) that can heat and eject gas from the galaxy, respectively. In our model, we only consider the latter effect that can eject gas out of the galactic environment. We assume each SNII to produce an energy equal to $E_{51}=10^{51}$erg of which a fraction ($f_w$) couples to the gas and drives the winds. In this work, we implement a ``delayed SN feedback" scheme \citep[see also][]{Mutch2016, Seiler2019} that accounts for the mass-dependent ($M_\mathrm{SN}$) lifetimes ($t_\mathrm{SN}$) of stars before they explode as SNII. We use the $M_\mathrm{SN}-t_\mathrm{SN}$ relation found by \citet{padovani1993} such that
\begin{eqnarray}
 t_\mathrm{SN} &=& \left[1.2\times10^3\ \left(\frac{M_\mathrm{SN}}{\msun}\right)^{-1.85} +\ 3\right]~\mathrm{Myr}.
 \label{eq_massSN}
\end{eqnarray}
In this case, stars of $M_\mathrm{SN} = 8~ (100)~\Msun$ explode as SNII $28.6 ~(3.23)$ Myr after star formation starts. We note that this $M_\mathrm{SN}-t_\mathrm{SN}$ relation may change in the presence of binaries \citep[see e.g.][]{zapartas2017}. However, given that the lifetimes of massive stars in binaries depend sensitively on the highly uncertain initial orbital period distribution and binary fraction (and the IMF and metallicity), we also maintain equation \ref{eq_massSN} for binaries as a conservative estimate. {\sc astraeus}, as many other semi-analytic galaxy evolution models, is based on merger trees that are discrete in time. Hence, the length of the time steps is key when deriving the total SNII energy. In case a time step exceeds $30$~Myrs, all high mass stars formed explode as SNII within that time step and SN feedback is ``instantaneous". However, as the time steps become increasingly shorter than $30$~Myrs, SNII feedback spans over multiple time steps, i.e., is ``delayed". The instantaneous SNII feedback scenario can therefore be regarded as a special case of the more general delayed SNII feedback.

We now describe our formalism for delayed SNII feedback. At any given redshift, the SNII energy that can couple to gas can be expressed as
\begin{eqnarray}
 E_\mathrm{SN}(z) &= & f_w E_{51} \left[ \sum_{j=1}^{N_j-1} \nu_j M_{\star,p}^\mathrm{new}(z_j) + \nu_z M_\star^\mathrm{new}(z) \right]. 
 \label{eq_energySN_discretised}
\end{eqnarray}
Here, for a given halo at redshift $z$, the first term on the right hand side represents the SNII explosions at $z$ from all the stars that formed in the progenitors of that halo, and the second term accounts for SNII from the newly formed stars at $z$. Further, $M_{\star,p}^\mathrm{new}(z_j)$ and $\nu_j$ are the newly formed stellar mass and the fraction of stars that explode as SN in time step $j$, between $z_j$ and $z_{j-1}$, respectively and $N_j$ is the number of simulation snapshots until and including $z$. Finally, $\nu_z$ is the fraction of the newly formed stars in the current time-step, $M_\star^\mathrm{new}(z)$, that explode as SNII. Using the assumed Salpeter IMF (with a slope of $\gamma=2.35$), the fraction of stars formed in the time interval $[t(z_{j-1}), t(z_j)]$ that explode as supernovae at step $z$ can be calculated as
\begin{eqnarray}
 \nu_j &=& \frac{2-\gamma}{1-\gamma}\ \frac{M_\mathrm{SN,j-1}^{1-\gamma} - M_\mathrm{SN,j}^{1-\gamma}}{M_\mathrm{star,low}^{2-\gamma} - M_\mathrm{star,high}^{2-\gamma}}.
\end{eqnarray}
Here $M_\mathrm{SN,j}$ is the mass of stars that would explode as SN after $t_\mathrm{SN}=t(z)-t(z_j)$ according to equation \ref{eq_massSN} with $t\geq t_j$, while $M_\mathrm{star,low}=0.1\msun$ and $M_\mathrm{star,high}=100\msun$.

In order to derive the amount of gas being ejected from the galaxy, we equate the SNII energy (see equation \ref{eq_energySN_discretised}) to the energy required to eject a gas mass (equal to $M_\mathrm{g}^\mathrm{ej}$) from a galaxy:
\begin{eqnarray}
 E_\mathrm{ej}(z) &=& \frac{1}{2} M_\mathrm{g}^\mathrm{ej}(z)\ v_e^2 = M_\mathrm{g}^\mathrm{ej}(z)\ v_c^2 = E_\mathrm{SN}(z),
 \label{eq_energyMgasEjected}
\end{eqnarray}
where $v_e$ is the ejection velocity that is related to the rotational velocity of the halo as $v_c = v_e/\sqrt{2}$.
This implies that the fraction of gas that needs to be converted into stars to eject the rest of the gas from the galaxy corresponds to
\begin{eqnarray}
 f_\star^{ej}(z) &=& \frac{M_\star^\mathrm{new}(z)}{M_\mathrm{g}^\mathrm{ej}(z) + M_\star^\mathrm{new}(z)} 
  \label{eq_fej_delayedSN} \\
 &=& \frac{v_c^2}{v_c^2 + f_w E_{51} \nu_z} \left[ 1 - \frac{f_w E_{51} \sum_{j=1}^{N_j-1} \nu_j M_{\star,j}^\mathrm{new}(z_j)}{M_\mathrm{g}^i(z)\ v_c^2} \right], \nonumber
\end{eqnarray}
where at maximum all the gas that is left after star formation can be ejected.

In case of instantaneous SN feedback equations \ref{eq_energyMgasEjected} and \ref{eq_fej_delayedSN} simplify to
\begin{eqnarray}
 M_\mathrm{g}^\mathrm{ej}(z) &=& \frac{f_w E_{51}}{v_c^2} \nu_z M_\star^\mathrm{new}(z),
 \label{eq_MgasEjected_instantaneousSN} \\
 f_\star^{ej}(z) &=& \frac{v_c^2}{v_c^2 + f_w E_{51} \nu_z},
 \label{eq_fej_instantaneousSN}
\end{eqnarray}
with $\nu_z=0.0077$~$\Msun^{-1}$ for the assumed Salpeter IMF.

In the remainder of this paper we use the delayed SN feedback scheme. As noted,  the snapshots of the N-body simulation scale as the logarithm of the scale-factor, resulting in increasingly longer time-steps with decreasing redshift $z$. Hence, while at $z\gtrsim9$ the delayed SN feedback scheme differs significantly from the instantaneous one, these schemes become increasingly similar with decreasing $z$ until there is effectively no difference at $z\lsim 6$.

\subsubsection{Resulting output of UV and \HI ionizing photons}
\label{subsubsec_emissivities}
We calculate the spectrum of each galaxy, $\xi(\nu,t)$, by convolving its star formation history with the starburst spectrum, $\xi_\mathrm{SP}(\nu,t)$,  obtained from two stellar population synthesis models (SPS): {\sc starburst99} \citep{leitherer1999} and {\sc bpass} that accounts for binaries \citep{eldridge2017}.
\begin{eqnarray}
 \xi(z) &=& \int_{\infty}^z \mathrm{d}z' \frac{\mathrm{d}t}{\mathrm{d}z'}\ \scalebox{0.93}{$\xi_\mathrm{SP}(\nu, t(z) - t(z'))\ M_\star^\mathrm{new}(z')\ f_\mathrm{lin}(\nu, z, z') $} \nonumber \\
 &=& \sum_\mathrm{j=1}^\mathrm{N_j} \xi_\mathrm{SP}(\nu, t-t_j)\ M_\star^\mathrm{new}(t_j)\ f_\mathrm{lin}(\nu, t, t_j),
 \label{eq_spectrum}
\end{eqnarray}
where $M_\star^\mathrm{new}(z')$ (or $M_\star^\mathrm{new}(t_j)$) is the newly formed stellar mass at redshift $z'$ (time step $t_j$). This newly formed stellar mass is assumed to form from star formation uniformly distributed over the entire time step\footnote{We note that we would yield similar emissivities, if we assumed a bursty star formation (being in agreement with our SN feedback scheme) within the time step.}. The factor $f_\mathrm{lin}$ accounts for this and is calculated as
\begin{eqnarray}
 f_\mathrm{lin}(\nu, t, t_j) &=& \frac{\int_{t_j}^{t_{j-1}} \mathrm{d}t'\ \xi_\mathrm{SP}(\nu, t-t')}{\xi_\mathrm{SP}(\nu, t_j)\ (t_{j}-t_{j-1})}.
\end{eqnarray}
where $t_{j-1}$ and $t_{j}$ are the beginning and end times of the time step with $t\geq t_{j}$.

The intrinsic spectrum of a stellar population sensitively depends on its age ($t$) and metallicity ($Z$). In the interest of simplicity, in this paper, we assume all stellar populations to have a stellar metallicity of $Z=0.05~\Zsun$ \citep{maio2010} and defer a full metallicity calculation to a future paper. In this paper, we focus on two key spectral quantities: (1) the number of \HI ionizing photons ($\lambda<912$ \AA~ in the rest-frame) that are required to understand the reionization of the IGM, and (2) the UV luminosity (rest-frame $1250-1500$\AA) to validate our model against observed Lyman Break Galaxy (LBG) data.

The intrinsic UV luminosity, $L_{\nu} [\mathrm{erg~s^{-1}~Hz^{-1}~\Msun^{-1}}]$, is quite similar in both the {\sc starburst99} and {\sc bpass} models and from the point in time when the stellar population was formed it evolves with time as
\begin{equation}
\frac{L_{\nu}(t)}{\scalebox{0.8}{$\mathrm{erg~s^{-1}~Hz^{-1}~\Msun^{-1}}$}} ~= 
 \begin{cases}
  8.24\times10^{20}&\mathrm{for}\ \frac{t}{\mathrm{Myr}}<4 \\
  2.07\times10^{21} \left[\frac{t}{2~\mathrm{Myr}} \right]^{-1.33} &\mathrm{for}\ \frac{t}{\mathrm{Myr}}\geq4.
 \end{cases} \\
 \label{eq_LUV}
\end{equation}

In contrast, the production rate of \HI ionizing photons, $\dot Q [\mathrm{s^{-1}~\Msun^{-1}}]$, sensitively depends on the SPS model used. In the {\sc starburst99} model, it evolves as 
\begin{equation}
 \frac{\dot Q(t)}{\mathrm{s}^{-1} \msun^{-1}} ~=~ 
 \begin{cases}
  3.63\times10^{46} &\mathrm{for}\ \frac{t}{\mathrm{Myr}}\leq3.16 \\
  2.18\times10^{47} \left[\frac{t}{2~\mathrm{Myr}} \right]^{-3.92} &\mathrm{for}\ \frac{t}{\mathrm{Myr}}>3.16,
 \end{cases} \\
 \label{eq_QSP_starburst99}
\end{equation}
while this quantity shows a shallower time-evolution in the {\sc bpass} model where 
\begin{equation}
 \frac{\dot Q(t)}{\mathrm{s}^{-1} \msun^{-1}} ~=~ 
 \begin{cases}
 3.20\times10^{46}&\mathrm{for}\ \frac{t}{\mathrm{Myr}}\leq3.16  \\
 9.09\times10^{46} \left[\frac{t}{2~\mathrm{Myr}} \right]^{-2.28} &\mathrm{for}\ \frac{t}{\mathrm{Myr}}>3.16.
 \end{cases} \\
 \label{eq_QSP_bpass}
\end{equation}

The total UV luminosity or ionizing photon output over any star formation history can be derived by using $\xi_\mathrm{SP} = L_{\nu}$ and $\xi_\mathrm{SP} = \dot Q$ in equation \ref{eq_spectrum}, respectively. 

\subsection{The reionization model}
\label{subsec_reion_model}
Most of the ionizing photons produced by a source (calculated from its star formation history as explained above) are absorbed within the interstellar medium with only a fraction ($f_\mathrm{esc}$) escaping and ionizing the IGM. This escaping rate of ionizing photons (the ionizing emissivity) can be expressed as 
\begin{eqnarray}
 \dot{N}_\mathrm{ion}(z) &=& f_\mathrm{esc}\ \dot Q(z),
\end{eqnarray}
with $\dot Q(z)$ being given by combining equations \ref{eq_spectrum} and \ref{eq_QSP_starburst99} (\ref{eq_QSP_bpass}) for {\sc starburst99} ({\sc bpass}).

For the majority of reionization scenarios that we consider in this paper, the ionizing escape fraction $f_\mathrm{esc}$ is assumed to be constant for all galaxies at all redshifts. However, we also explore a scenario where $f_\mathrm{esc}$ is coupled to the fraction of gas that is ejected from the galaxy into the IGM. This is supported by a number of simulations that find that SN explosions create under-densities through which ionizing photons can escape \citep{Wise2009, Wise2014, Kimm2014, Kimm2017, Xu2016}, implying that the ionizing escape fraction $f_\mathrm{esc}$ increases as a larger fraction of gas is pushed into outflows. In this case we model the ionizing escape fraction $f_\mathrm{esc}$ as,
\begin{eqnarray}
 f_\mathrm{esc} &=& f_\mathrm{esc}^0\ \frac{\feff}{f_\star^\mathrm{ej}}.
 \label{eq_fesc}
\end{eqnarray}
where $f_\mathrm{esc}^0$ is a free parameter that can be tuned to adjust the timing of reionization\footnote{For instantaneous SN feedback, equation \ref{eq_fesc} can be expressed analytically by $f_\mathrm{esc}^0 \times \min\left[1, f_\star \left( 1 + \frac{f_w E_{51} \nu_z}{\left(3\pi H_0\right)^{2/3} \Omega_m^{1/3} (1+z) M_h^{2/3}} \right) \right]$.}. This ansatz results in a very high escape fractions for low-mass ($\lsim 10^{9.5}\Msun$) galaxies where $\feff=f_\star^\mathrm{ej}$. As the gravitational potential of the galaxy deepens, the SN explosions of the stars that are forming can not eject all gas from the galaxy ($\feff\ll f_\star^\mathrm{ej}$) and $f_\mathrm{esc}$ drops down to a few percent for $M_h\sim10^{11}\Msun$ halos  and assuming $f_\mathrm{esc}^0=1$.

These escaping ionizing photons both reionize and heat the IGM. The amount of heating, naturally, critically depends on the energy of the ionizing photons (i.e. the hardness of the source spectrum). For star-forming galaxies the ionized IGM can be heated up to $\sim10^4$~K \citep[e.g.][]{schaye2000}, which has two important effects: firstly, as the gas residing in halos heats up, the higher pressure causes a fraction of it to photo-evaporate into the IGM, reducing the amount available for star formation \citep{barkana-loeb1999, shapiro2004}. Secondly, since a higher IGM temperature corresponds to a higher Jeans mass, the minimum mass for galaxy formation increases, thereby reducing the amount of gas being accreted by the galaxy \citep{couchman-rees1986, efstathiou1992, Hoeft2006}. These mechanisms lead to a reduction of gas mass, particularly in low-mass halos where the gravitational potential is not deep enough to compensate for the increased pressure of the heated gas. While the rise in gas temperature occurs quasi instantaneously, the gas pressure adjusts over the dynamical time scale of the galaxy \citep{Gnedin2000}, which leads to a time delay between the time of reionization and the onset of the gas (and star formation) suppressing effect of radiative feedback.
However, modelling these radiative feedback processes remains challenging due to their complex dependence on the halo mass and redshift, the patchiness and strength of the UVB and the redshift at which an assembling halo is first irradiated by the UV background \citep{Gnedin2000, sobacchi2013a}.

In this work, we explore a wide range of physically plausible radiative feedback models in order to study their impact on both the galaxy population and progress of reionization whilst ensuring agreement with all available (galaxy and reionization) data sets. It is essential to account for the patchiness of reionization: for example, the impact of radiative feedback might be more severe on galaxies forming in an over-dense region reionized early-on as compared to those forming in an under-dense region reionized later. As noted before, in order to simulate the galaxy formation-reionization interplay, we couple galaxy formation (simulated through the {\sc delphi} semi-analytic galaxy evolution model) with a semi-numerical reionization code \citep[{\sc cifog;}\footnote{\url{https://github.com/annehutter/grid-model}}][]{Hutter2018} in a self-consistent manner.

{\sc cifog} is a MPI-parallelised, semi-numerical reionization code that computes the time- and spatial-evolution of ionized regions in the IGM. Here we provide a brief overview and refer the interested reader to \citet{Hutter2018} for details. Essentially, {\sc cifog} follows the approach outlined in \citet{Furlanetto2004} where a spherical region is considered to be ionized if the cumulative number of ionizing photons ($N_\mathrm{ion}$) emitted exceeds the cumulative number of absorption events ($N_\mathrm{abs}$), and neutral otherwise. Starting with large radii and decreasing the size of the region by reducing the sphere radius $R$, the central cell of the spherical region is considered ionized if
\begin{eqnarray}
 N_\mathrm{ion}(z) &=& \sum_{i=0}^{N_\mathrm{gal}(R)} \left[ \int_z^{\infty} \mathrm{d}z' \frac{\mathrm{d}t}{\mathrm{d}z'}\ \langle\dot{N}_\mathrm{ion,i}\rangle_R(z') \right] \\
 \geq && \nonumber \\
 N_\mathrm{abs}(z) &=& \langle n_\mathrm{H,0} \rangle_R V_\mathrm{cell} \left[ 1 + \int_z^{z_\mathrm{reion}} \mathrm{d}z' \frac{\mathrm{d}t}{\mathrm{d}z'}\ \langle \dot{N}_\mathrm{rec} \rangle_R(z') \right]. \nonumber
\end{eqnarray}
Here, $\dot{N}_\mathrm{ion}(z)$ is the ionizing emissivity of a galaxy $i$ at redshift $z$ located within the sphere of radius $R$ and $N_\mathrm{gal}(R)$ is the number of galaxies within that sphere; $\langle \rangle_R$ indicates that the quantity is averaged over a sphere with radius $R$. Further, $n_\mathrm{H,0}$, $V_\mathrm{cell}$ and $\dot{N}_\mathrm{rec}(z)$ are the hydrogen density at $z=0$, the comoving volume of the cell and the recombination rate at $z$, respectively.  Applying the ionization criterion in large enough regions ensures that the radiation from neighbouring sources is accounted for.

{\sc cifog} derives the residual \HI fraction and recombination rate of each cell from the local gas density and photoionization rate. The code supports two models for the spatially-dependent photoionization rate: one that is based on the mean free path given by the size of the ionized regions (mean free path approach), and one that is based on the flux of the ionizing sources (flux based approach). In this work we utilise the flux based approach\footnote{We note that {\sc astraeus} supports all features that {\sc cifog} offers, i.e. different photoionization models or flagging regions as ionized (central cell versus entire sphere) can be chosen.}.

In this work, we use the density fields that have been obtained by mapping the DM particles on to a $2048^3$ grid using the cloud-in-cell (CIC) algorithm. {\sc cifog} then runs on $512^3$ grids that have been obtained by reducing the $2048^3$ grids. Furthermore, we note that the time and spatial evolution of the larger ionized regions computed with {\sc cifog} are hardly affected by the resolution of the grid. However, ionized regions smaller than a grid cell are not spatially resolved, which leads to the cell being reionized when the volume of the cell is ionized. A better resolution would resolve those ionized regions and provide more accurate times of reionization ($z_\mathrm{reion}$), particularly around low-mass galaxies.

{\sc delphi} and {\sc cifog} are coupled in a self-consistent manner using the following approach at each time step:
\begin{enumerate}
 \item {\sc delphi} evolves galaxies from $z_{j-1}$ to $z_j$ and computes the {\it ionizing emissivity} of each galaxy at $z_j$ from the star formation histories that it stores for all galaxies.
 \item At each redshift, the ionizing emissivities of the galaxies are fed into {\sc cifog}. Accounting for the location of the galaxies in the simulated large-scale structure, {\sc cifog} computes the {\it time evolution of the ionized regions} in the IGM on a $512^3$ grid from $z_j$ to $z_{j+1}$.
 \item In the subsequent time step ($z_{j+1}$), we identify each galaxy whose cell was reionized in previous time steps $z>z_{j+1}$. For galaxies lying in reionized regions, we track their redshift of reionization $z_\mathrm{reion}$ and the incident photoionization rate at $z_\mathrm{reion}$, and calculate the fraction of gas mass they can retain after radiative feedback ($f_g$) using the prescriptions detailed in Sec. \ref{subsec_Mc} that follows; galaxies in neutral regions are naturally unaffected by reionization feedback. Accounting for radiative feedback, {\sc delphi} evolves all galaxies from $z_{j}$ to $z_{j+1}$.
\end{enumerate}

\subsection{Resulting characteristic masses of supernova and radiative feedback processes}
\label{subsec_Mc}
As described above, both supernova and radiative feedback affect the gas content of galaxies, with the feedback efficiency generally decreasing as the gravitational potential of the host halo increases. This results in a ``characteristic" mass that can be associated with each feedback process - this is defined as the mass at which the halo can still hold on to some and half of its original gas mass for supernova and radiative feedback, respectively. We now discuss the characteristic masses for SN feedback and the different radiative feedback models that have been implemented in {\sc astraeus} and are summarised in Table \ref{tab_best_fit_values}. 

\begin{figure*}
 \centering
 \includegraphics[width=1.0\textwidth]{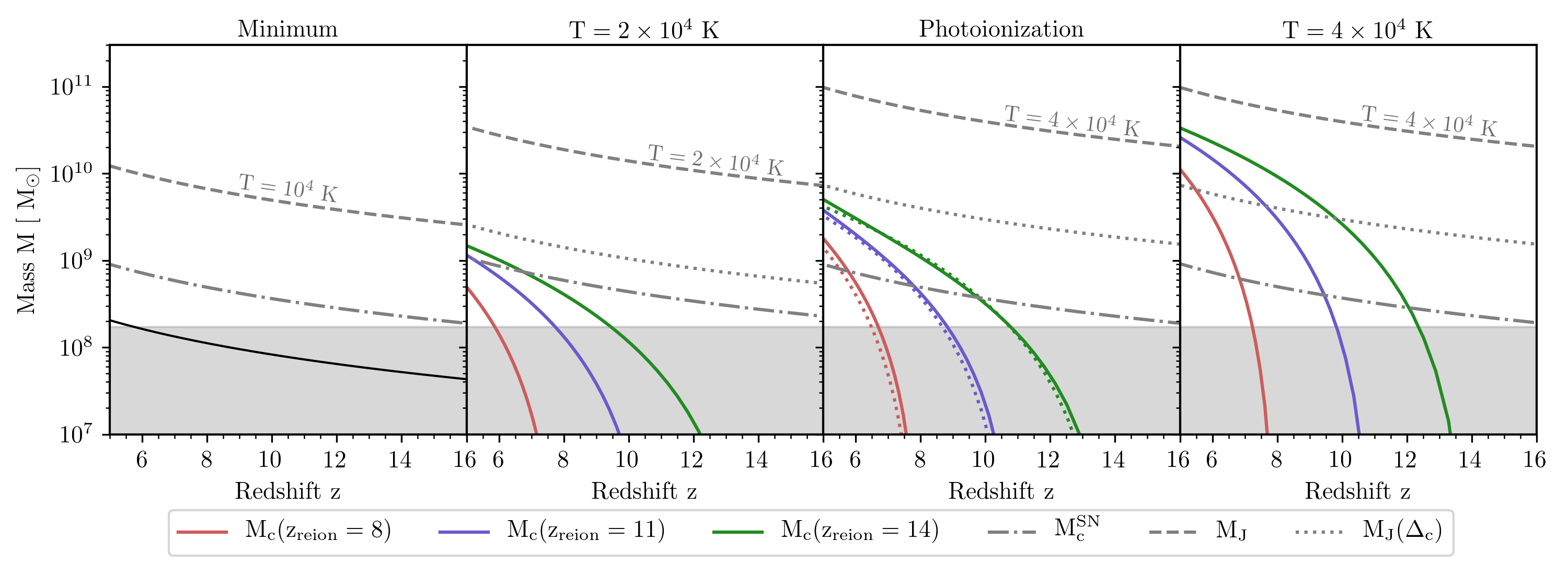}
 \caption{Characteristic masses for the radiative and SN feedback processes ($M_c$ and $M_c^\mathrm{SN}$) as well as the Jeans mass ($M_J$). From left to right the masses are shown for our radiative feedback models {\it Minimum}, {\it Heating} with $T_0=2\times10^4$K and $M_c=M_F$, {\it Photoionization} with $\Gamma_\mathrm{HI}=10^{-12.3}$s$^{-1}$, and {\it Heating} with $T_0=4\times10^4$~K and $M_c=8M_F$. Red, blue and green solid lines correspond to the characteristic masses due radiative feedback when the region has been reionized at $z=8$, $11$ and $14$, respectively. In the {\it Minimum} model the filtering mass is independent of reionization and is shown by the black solid line. In the third panel, coloured, dotted lines show the characteristic masses for a {\it Heating} model with $T_0=4\times10^4$~K and $M_c=M_F$. Grey, dash-dotted lines show the characteristic masses for SN feedback, while grey dashed and dash-dotted lines the Jeans mass at mean density and virial over-density for the temperature indicated. The grey shaded area marks the halo masses that are not resolved in our simulation.}
 \label{fig_filtering_masses}
\end{figure*}

\subsubsection{Characteristic mass for supernova feedback}
\label{subsubsec_Mc_SN}
Given that SN feedback can eject gas from every star forming galaxy, we define its characteristic mass ($M_c^\mathrm{SN}$) as the minimum halo mass which can hold onto a non-zero value of its gas mass after star formation. In the case of instantaneous SN feedback, this can be calculated, by equating the supernova energy coupling to gas to the binding energy of the gas left over after star formation, as
\begin{eqnarray}
 M_c^\mathrm{SN} = 
 5.6\times10^{13}\msun \left( \frac{f_\star f_w}{1-f_\star} \right)^{3/2} \Omega_m^{-1/2} \left(1+z\right)^{-3/2} .
 \label{eq_filtering_mass_SN}
\end{eqnarray}
The redshift evolution of $M_c^\mathrm{SN}$ is shown as a reference in all panels by the dash-dotted lines in Fig. \ref{fig_filtering_masses}. As seen, SN feedback can eject all of the gas mass in halos less massive than $10^{8.3}\Msun$ at $z \sim 16$. As the matter density of the Universe decreases with cosmic time, the gravitational potential corresponding to a given halo mass becomes shallower, leading to an increase in the supernova characteristic mass to $M_c^\mathrm{SN} \sim 10^{9}\Msun$ by $z \sim 5$.

\subsubsection{Atomic cooling mass (Minimum feedback model)}
\label{subsubsec_Mc_min}
This is the {\it weakest} of our radiative feedback models. Here, ionized IGM gas is assumed to be heated to $T=10^4$~K via photo-heating. Only halos massive enough to have virial temperatures exceeding $10^4$~K can maintain all of their gas; lower mass halos are assumed to be completely gas-free. The gas fraction left after radiative feedback ($f_g$) is obtained by comparing the halo mass to the (cooling) mass within the virial radius at the critical over-density for collapse, $\Delta_c \rho_c\simeq18\pi^2\rho_c$ \citep[e.g.][]{Barkana2001} such that
\begin{eqnarray}
 M_\mathrm{cool}(z) &=& 4.5\times10^7h^{-1}\msun  \left(\frac{\Omega_m}{0.3} \right)^{-1/2} \left(\frac{1+z}{10}\right)^{-3/2} \nonumber \\
 && \left(\frac{\mu}{0.6}\right)^{-3/2} \left(\frac{T_\mathrm{vir}}{10^4\mathrm{K}} \right)^{3/2}.
\end{eqnarray}
using $T_\mathrm{vir}=10^4$ K. Then $f_g$ is calculated to be
\begin{eqnarray}
 f_g &=&
 \begin{cases}
  0 & \quad \text{if } M_\mathrm{vir}(z)<M_\mathrm{cool}(z) \\
  1 & \quad \text{if } M_\mathrm{vir}(z)\geq M_\mathrm{cool}(z).
 \end{cases}
\end{eqnarray}
From the first panel in Fig. \ref{fig_filtering_masses}, we see that the characteristic mass for SN feedback always exceeds that for the atomic cooling mass. As a result, our {\it Minimum} radiative feedback model shows no impact of radiative feedback on either galaxy formation or reionization. 

\subsubsection{Temperature-dependent characteristic masses (Heating models)}
\label{subsubsec_Mc_heating}
\citet{gnedin1998b} introduced a filtering scale, $k_F$, below which baryon fluctuations are suppressed as a consequence of reionization heating. This filtering scale can be linked to a filtering mass \citep{Gnedin2000, Okamoto2008, Naoz2013}
\begin{eqnarray}
 M_F &=& \frac{4\pi}{3} \overline{\rho} \left( \frac{\pi a}{k_F} \right)^3,
\end{eqnarray}
where $\overline{\rho}$ is the average total mass density of the Universe and $a$ the scale factor. Applying the filtering scale for baryons to the continuity equation that describes the linear evolution of perturbations in the dark matter-baryon fluid, it has been shown that whilst the filtering scale is related to the Jeans scale as a function of time, at a given time those two scales are unrelated \citep{gnedin1998b}; determining $k_F$ at a given time therefore requires knowing the evolution of the Jeans scale up to that time. This is due to the fact that the response of the gas density distribution to an immediate temperature change occurs on the dynamical timescale.  The filtering mass can be calculated as (at $z>2$)
\begin{eqnarray}
 M_F^{2/3} &=& M_{J_0}^{2/3} \frac{3}{a} \int_0^a \mathrm{d}a'\ a'\ T_4(a') \left[ 1-\left( \frac{a'}{a} \right)^{1/2} \right],
 \label{eq_filtering_mass}
\end{eqnarray}
where $M_{J_0}$ is the Jeans mass at $T=10^4$~K and $z=0$, and $T_4(a)$ is the evolving baryonic temperature in units of $10^4$~K. 
The Jeans mass at redshift $z$ depends on the Jeans scale $k_J=a ~ c_s^{-1} (4\pi G \overline{\rho})^{1/2}$ and the linear-theory sound speed $c_s=\sqrt{5 k_B T [3 \mu m_p]^{-1}}$, and can be calculated as
\begin{eqnarray}
 M_J(z) &=& \frac{4\pi}{3} \overline{\rho} \left( \frac{\pi a}{k_J} \right)^3
  \label{eq_jeans_mass}\\
 &=& \frac{3.13\times10^{10} h^{-1} \msun}{\Omega_m^{1/2} \left( 1+z\right)^{3/2}} \mu^{-3/2} \left(\frac{T}{10^4\mathrm{K}}\right)^{3/2}. \nonumber
\end{eqnarray}
Furthermore, we model the redshift evolution of the baryonic temperature as
\begin{eqnarray}
 T_4(a) &=&
 \begin{cases}
  \frac{T_\mathrm{CMB}}{10^4\mathrm{K}}\ \frac{1}{a}, & \quad \text{if } a_\mathrm{rec} \leq a < a_\mathrm{dec} \\
  \frac{T_\mathrm{CMB}}{10^4\mathrm{K}}\ \frac{a_\mathrm{dec}}{a^2}, & \quad \text{if } a_\mathrm{dec} \leq a < a_\mathrm{reion} \\
  \frac{T_0}{10^4\mathrm{K}}\ \left( \frac{a}{a_\mathrm{reion}} \right)^{-1}, & \quad \text{if } a_\mathrm{reion} \leq a.
 \end{cases}
 \label{eq_temperature_evolution}
\end{eqnarray}
These three terms correspond to the epoch after recombination ($a_\mathrm{rec}=1/1100$) where gas is still coupled to the cosmic background radiation by Compton heating, the epoch after decoupling ($a_\mathrm{dec}=1/251$) when gas cools adiabatically, and the epoch of reionization and subsequent cooling \citep{hui1997}, respectively. The IGM temperature, $T_0$, is a free parameter. 

\begin{table*}
\begin{threeparttable}
 \begin{tabular}{|c|c|c|c|c|c|c|}
 \hline
  & {\it Minimum} & {\it Weak Heating} & {\it Photoionization} & {\it Early Heating} & {\it Strong Heating} & {\it Jeans Mass} \\
  \hline
  \hline
  $f_\star$ & 0.01 & 0.01 & 0.01 & 0.01 & 0.011 & 0.01 \\
  \hline
  $f_w$ & 0.2 & 0.2 & 0.2 & 0.2 & 0.19 & 0.2 \\
  \hline
  $f_\mathrm{esc}^\mathrm{S99}$ & 0.21 & 0.21 & 0.215 & 0.60\footnotemark[1] & 0.22 & 0.285 \\ 
  \hline
  $f_\mathrm{esc}^\mathrm{BPASS}$ & 0.0185 & 0.0185 & 0.019 & 0.052 & 0.019 & 0.025 \\ 
  \hline
  $M_c$ & $M_\mathrm{cool}(z, T)$ & $M_F(z, z_\mathrm{reion}, T)$ & $M_c(z, z_\mathrm{reion}, \Gamma_\mathrm{HI})$ & $M_F(z, z_\mathrm{reion}, T)$ & $8 M_F(z, z_\mathrm{reion}, T)$ & $M_J(z, T)$ \\
  \hline
  $T_0$ & $10^4\mathrm{K}$ & $2\times10^4\mathrm{K}$ & - & $2\times10^4\mathrm{K}$ & $4\times10^4\mathrm{K}$ & $4\times10^4\mathrm{K}$ \\
  \hline
 \end{tabular}
 \begin{tablenotes}
  \item[1]{\scriptsize {This value represents $f_\mathrm{esc}^0$ and is the maximum that $f_\mathrm{esc}$ in the {\it Early Heating} model can adopt.}}
 \end{tablenotes}
 \end{threeparttable}
 \caption{For the different radiative feedback scenarios considered in this work (shown by the different columns) we show the parameter values for the threshold star formation efficiency ($f_\star$), the fraction of SNII energy coupling to gas ($f_w$), the escape fraction for ionizing photons for the {\sc Starburst99} and {\sc BPASS} stellar population synthesis models ($f_\mathrm{esc}^\mathrm{S99}$ and $f_\mathrm{esc}^\mathrm{BPASS}$ respectively), the characteristic mass for radiative feedback ($M_c$) and the IGM temperature in ionized regions ($T_0$). Further, $f_\star$, $f_w$ and $f_\mathrm{esc}$ ($f_\mathrm{esc}^\mathrm{S99}$ or $f_\mathrm{esc}^\mathrm{BPASS}$ for {\sc starburst99} or {\sc bpass}, respectively) are our model free parameters that are tuned to simultaneously reproduce all high-redshift galaxy and reionization data sets. These model parameters have similar and even identical values, since our radiative feedback models affect only low-mass and faint galaxies where observational constraints are sparse. Extreme models that alter the ionizing emissivities of galaxies, either through suppression of star formation ({\it Jeans Mass}) or an $f_\mathrm{esc}$ depending on the fraction of gas ejected from the galaxy ({\it Early Heating}) show higher $f_\mathrm{esc}$ values.}
 \label{tab_best_fit_values}
\end{table*}

Although a number of hydrodynamical simulations of galaxy populations with inhomogeneous or homogeneous UV backgrounds \citep{Gnedin2000, Okamoto2008, Naoz2013} have shown that $M_F$ can be related to the characteristic mass $M_c$ (the halo mass that on average retains $50\%$ of its gas mass), the exact relation remains debated: while \citet{Gnedin2000} obtain $M_c\simeq8 M_F$, other works yield $M_c\simeq M_F$ \citep{Naoz2013}.
From the characteristic mass, the gas fraction maintained by the galaxy can be found  following the fitting formula provided in \citet{Gnedin2000} such that
\begin{eqnarray}
 f_g &=& \left[ 1 + (2^{1/3}-1)\frac{M_c}{M_\mathrm{vir}} \right]^{-3}.
\end{eqnarray}

In this paper, we explore three scenarios for such a {\it Heating} model: the weakest ({\it Weak Heating}) and the strongest ({\it Strong Heating}) have been chosen to bracket the physically plausible range of radiative feedback for this model:

{\it (i) Weak Heating}: Here we assume the reionized IGM is heated to $T_0=2\times10^4$~K and the characteristic mass for radiative feedback is equal to the filtering mass, i.e., $M_c=M_F$. From the second panel in Fig. \ref{fig_filtering_masses}, we can see that only galaxies reionized very early-on (i.e. at $z\gtrsim14$) are affected more by radiative feedback as compared to SN feedback. Galaxies reionized later on are only affected if their halo masses are less than $\sim10^{8-9} \Msun$. 

{\it (ii) Early Heating}: In this model we use an IGM temperature of $T_0=4\times10^4$~K and $M_c=M_F$, resulting in similar characteristic masses as the {\it Photoionization} model described in the next Section. However, this model is designed to explore the extent to which the impact of radiative feedback can be enhanced for low-mass galaxies ($M_h\lesssim10^{9.5}\msun$) by reionizing them earlier. In order to maximise this effect, whilst remaining in agreement with the {\it Planck} optical depth measurements, we assume the ionizing escape fraction $f_\mathrm{esc}$ to scale with the ejected gas fraction, resulting in a decreasing $f_\mathrm{esc}$ with halo mass. For identical $f_\mathrm{esc}$, $f_\star$ and $f_w$, such a {\it Heating} model produces very similar results to the {\it Photoionization} model discussed below (c.f. dotted to solid coloured lines in third panel in Fig. \ref{fig_filtering_masses}). 

{\it (iii) Strong Heating}: As in the {\it Early Heating} model, we assume an IGM temperature of $T_0=4\times10^4$~K. However, in order to increase the impact of radiative feedback, we assume the radiative feedback characteristic mass to be $8$ times the filtering mass, i.e., $M_c = 8M_F$. From the last panel in Fig. \ref{fig_filtering_masses}, we see that even galaxies reionized later (e.g. $z\lesssim8$) exceed the characteristic mass for SN feedback. Indeed, at $z\simeq6$, even galaxies with halo masses up to $\sim10^{10} \Msun$ show suppression of star formation in this model. From the point in time, when a galaxy becomes reionized, radiative feedback dominates over SN feedback for a time period corresponding to $\Delta z\simeq1-1.5$. 

\subsubsection{Photoionization rate dependent characteristic mass (Photoionization model)}
\label{subsubsec_Mc_Sobacchi}
Using 1D radiation hydrodynamical simulations and assuming an IGM temperature of $T=0 ~ (10^4)$~K in neutral (ionized) regions,
\begin{eqnarray}
 T({\bf x})&=&
 \begin{cases}
 0~\mathrm{K} & \quad \text{if } \chi_\mathrm{HII}({\bf x}) = 0 \\
 10^4~\mathrm{K} & \quad \text{if } \chi_\mathrm{HII}({\bf x}) = 1
 \end{cases}
 \label{eq_temperature_sobacchi}
\end{eqnarray}
\citet{sobacchi2013a} have derived the following ansatz for the critical mass using the filtering mass approach proposed in \citet{Gnedin2000}'s:
\begin{eqnarray}
 M_c(J, z, z_\mathrm{reion}) &=& J_{21}^\alpha g_1(z) g_2(z, z_\mathrm{reion}).
\end{eqnarray}
This is motivated by the fact that, inserting their temperature relation (equation \ref{eq_temperature_sobacchi}) into equation \ref{eq_filtering_mass}, we can see that only $M_{J,0}$ (or $k_J$) is dependent on the temperature $T_0$, which again can be expressed in terms of the ionizing background $J_{21}$. Quantitatively, the critical mass is found to be \citep{sobacchi2013a, sobacchi2014}
\begin{eqnarray}
 M_c(M_0, a, b, c, d) = M_0 J_{21}^a \left( \frac{1+z}{10} \right)^b \left[ 1 - \left( \frac{1+z}{1+z_\mathrm{reion}} \right)^c \right]^d \nonumber
\end{eqnarray}
with best-fit values of $M_0=2.8\times10^9\Msun$, $a=0.17$, $b=-2.1$, $c=2$, $d=2.5$, and $J_{21}=(\Gamma_\mathrm{HI}/10^{-12})\mathrm{s}^{-1}$ where $\Gamma_\mathrm{HI}$ is the photoionization rate at the location of the galaxy at the time its environment was reionization. In this case, the the fraction of gas that is retained by a halo can be expressed as
\begin{eqnarray}
 f_g &=& 2^{-M_c/M_h}.
\end{eqnarray}
We note that the photoionization rate $\Gamma_\mathrm{HI}$ is a proxy for the IGM temperature. We find that a photoionization rate of $\Gamma_\mathrm{HI}=10^{-12.3}$s$^{-1}$ corresponds to a temperature of $T_0\simeq4\times10^4$~K in the {\it Heating} model.\footnote{The relation between $\Gamma_\mathrm{HI}$ and $T_0$ has been derived from analysing the $\Gamma_\mathrm{HI}$ values in over-dense cells at the time of reionization as well as comparing observables such as the UV luminosity and stellar mass functions.}

In this {\it Photoionization} model, as shown (by the solid coloured lines) in the third panel in Fig. \ref{fig_filtering_masses}, only galaxies reionized early-on, i.e. at $z\gsim10$, will experience sufficient radiative feedback. Galaxies reionized later are only affected by radiative feedback when they have a smaller gravitational potential, i.e. are less massive than $\sim10^9 \Msun$ at $z\simeq6$. The dotted coloured lines in the third panel in Fig. \ref{fig_filtering_masses} show results for the corresponding {\it Heating} model with an IGM temperature of $T_0=4\times10^4$~K and $M_c=M_F$.

\subsubsection{Jeans mass (Maximum feedback model)}
\label{subsubsec_Mc_jeans}
The strongest of our radiative feedback models is the Jeans mass model. Here, the gas in the IGM is assumed to be heated to $T_0 = 4\times10^4$~K via photoheating upon ionization. However, the rise in temperature is assumed to translate immediately into a lower gas density and hence a higher Jeans mass at the virial over-density. 
The fraction of gas $f_g$ that is maintained by a galaxy in an ionized region is given by
\begin{eqnarray}
 f_g &=& 2^{-M_J / M_h},
\end{eqnarray}
where $M_J$ is determined by equation \ref{eq_jeans_mass}.

This Jeans mass at each redshift is shown (as the dotted grey line) in the fourth panel in Fig. \ref{fig_filtering_masses}. As soon as the cell hosting a galaxy is reionized, galaxies with halo masses $M_h \lsim M_J$ are {\it immediately} affected by radiative feedback.
We have used this model in order to delimit the maximum possible effect of radiative feedback on early galaxy formation. In addition,  such a model is often employed in reionization simulations \citep[e.g.][]{iliev2007, iliev2012, Dixon2016} and therefore provides a useful comparison against previous works.

\begin{figure*}
 \centering
 \includegraphics[width=0.95\textwidth]{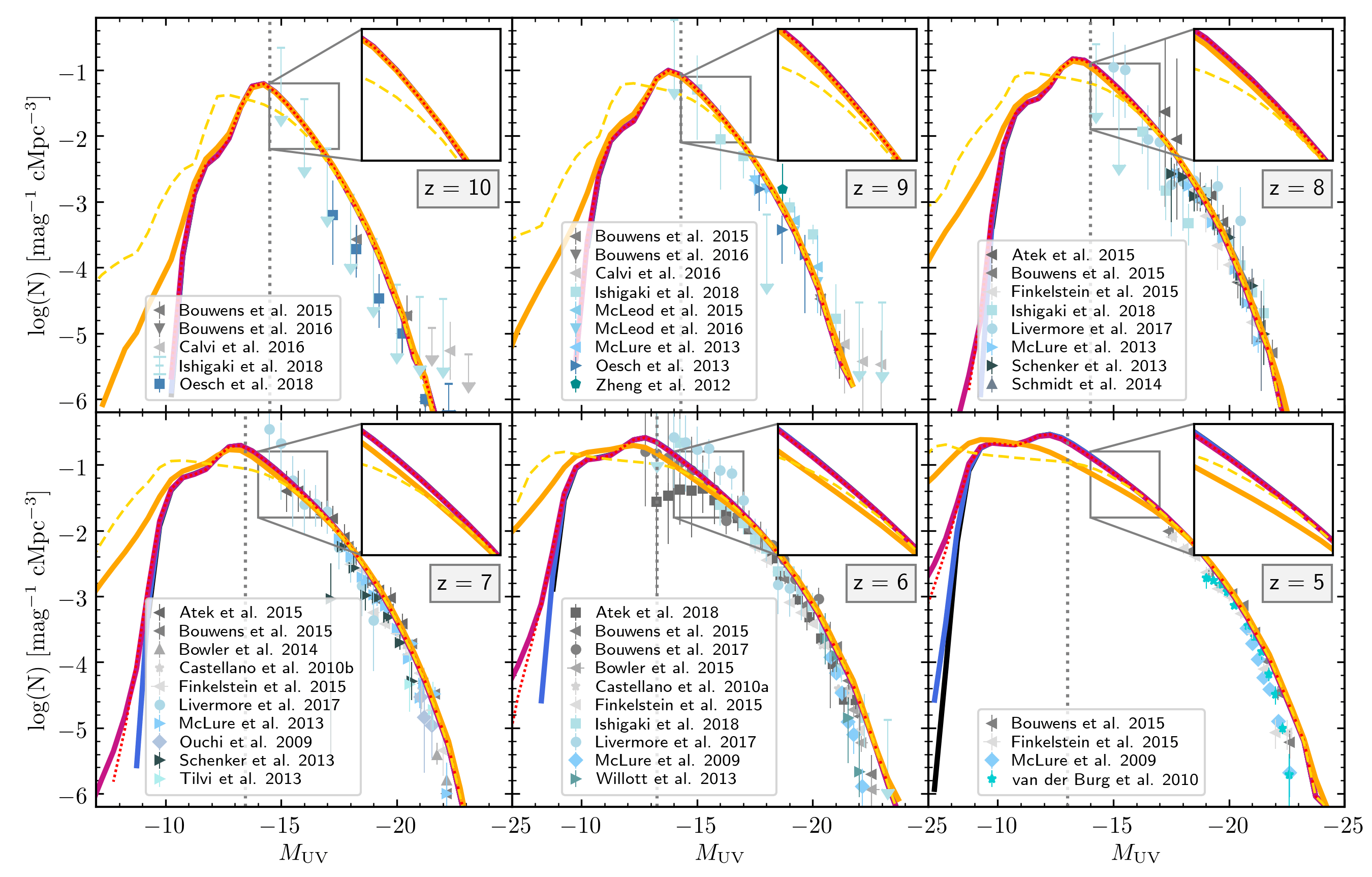}
 \caption{UV luminosity functions (UV LFs) at $z \sim 5-10$ using the best-fit parameters noted in Table \ref{tab_best_fit_values} and accounting for all galaxies in halos with $M_h\geq10^{8.6}\msun$. In each panel we show results for the different radiative feedback models studied in this work: {\it Minimum} (black solid line), {\it Weak Heating} (blue solid line), {\it Photoionization} (violet solid line), {\it Early Heating} (red dotted line), {\it Strong Heating} (orange solid line) and {\it Jeans Mass} (yellow dashed line). In each panel, the vertical grey dotted line indicates the UV luminosities of galaxies below which the UV LF is affected by the resolution of the underlying N-body simulation {\sc vsmdpl}. Finally, the grey and blue-grey points indicate observational data collected by different works, as marked in each panel \citep{Atek2015, Atek2018, Bouwens2015, Bouwens2016, Bouwens2017, Bowler2014, Bowler2015, Calvi2016, Castellano2010a, Castellano2010b, Finkelstein2015, Ishigaki2018, Livermore2017, McLeod2015, McLeod2016, McLure2009, McLure2013, Oesch2013, Oesch2018, Ouchi2009, Schenker2013, Schmidt2014, Tilvi2013, vanderBurg2010, Willott2013, Zheng2012}.}
 \label{fig_UV_LFs}
\end{figure*}

\section{Baselining the model against observed data-sets}
\label{sec_fitting}

We tune the three free mass- and redshift-independent parameters of our framework ($f_\star, ~f_w$ and $f_\mathrm{esc}$) for each radiative feedback model by simultaneously matching to a number of galaxy observables (the UV luminosity functions at $z=5-10$, the stellar mass functions at $z=5-10$, the redshift evolution of the stellar mass and star formation rate) and reionization data-sets (constraints on the ionization history inferred using quasars, Lyman-$\alpha$ emitters, Gamma Ray Bursts and the integrated electron scattering optical depth)\footnote{ In practise, we firstly adjust $f_\star$ and $f_w$ to reproduce primarily the UV LFs and SFRD evolution and then the SMFs and SMD evolution, and secondly tune $f_\mathrm{esc}$ to reproduce the integrated electron optical depth.}. The best fit values of the free parameters for each radiative feedback model are listed in Table \ref{tab_best_fit_values}. 
We note that our free parameters should be thought of as the ``observed" values, since we calibrate our model to observations without correcting for effects such as dust attenuation\footnote{Our model does not include dust at this stage. However, dust only plays a significant role in the attenuation of bright galaxies at $z\lesssim7$.}.

\subsection{Redshift evolution of the Ultra-violet luminosity function}
\label{subsec_UV_LF}

For all galaxies in our simulation, we calculate the UV luminosities at 1500\AA~ at $z=5-10$ from their entire star formation histories (SFH) by inserting equation \ref{eq_LUV} for $\xi_\mathrm{SP}$ into equation \ref{eq_spectrum}.
In Fig. \ref{fig_UV_LFs}, we show the UV luminosity functions (UV LFs) for all our radiative feedback models and accounting for all galaxies with a halo mass of $M_h\geq10^{8.6}\msun$. We start by noting that while our model results are in broad agreement with the observed UV LF at $z \sim 5-10$, they slightly over-predict the number density of bright galaxies at $z \lsim 6$. This is probably due to the fact that our model does not account for the increasing dust attenuation expected for massive galaxies with cosmic time.

In our model the bright end of the UV LF is determined by the threshold star formation efficiency $f_*$, while the faint end of the UV LF ($M_\mathrm{UV}\gtrsim-15$) is shaped by a combination of supernova ($f_w$) and radiative feedback ($M_c$). While the evolution of the bright end is driven by a genuine luminosity evolution as galaxies grow in (halo, gas and stellar) mass through mergers and gas accretion, the evolution of the faint end is more complex due to the stochastic star formation in low-mass halos induced by SN feedback and  by radiative feedback. Indeed, at the faint end, the UV LF evolution involves a combination of positive and negative luminosity evolution (as low-mass galaxies brighten and fade) and a positive and negative density evolution (as new low-mass galaxies form are consumed by merging) as pointed out in previous works \citep[e.g][]{dayal2013, yung2019}.

In order to assess the role of SN feedback in shaping the faint-end of the UV LF, we start by discussing the key features of the UV LFs in our {\it Minimum} radiative feedback model (see black solid lines in Fig. \ref{fig_UV_LFs}, at most redshifts overlapped by the blue ({\it Weak Heating}) and violet lines ({\it Photoionization})). In this model, the characteristic mass for radiative feedback only exceeds the halo resolution mass ($1.2\times10^8h^{-1}\Msun$) at $z \lsim 5.8$ (see Fig. \ref{fig_filtering_masses}) and always remains below the characteristic mass for SN feedback. This results in SN feedback dominating over radiative feedback at all redshifts in this model. As cosmic time proceeds and the density contrast in the Universe increases, the turn over or peak of the UV LF broadens and shifts to fainter UV luminosities.
In order to explain these trends, we start by examining the characteristic UV luminosity of galaxies in newly-formed halos at the resolution limit of the simulation ($M_h\simeq1.2\times10^8h^{-1}\Msun$). 
Initially, these galaxies are gas-rich and have a burst of star formation with the exact UV luminosity depending on the SN feedback efficiency ($f_w$) and redshift;  e.g. newly formed halos have a UV magnitude corresponding to $M_\mathrm{UV}\simeq-14$ at $z\simeq10$ which increases to $M_\mathrm{UV}\simeq-12.5$ by $z\simeq5$). The complete loss of (SN-driven) gas mass, which can not be compensated by accretion in a subsequent time step, results in almost no new star formation after the initial burst. This results in a continual decrease in the UV luminosity of such low-mass galaxies. Indeed, as such galaxies age, the UV luminosity only drops. As a result, fainter UV luminosities can be reached in gas-poor low-mass galaxies as they age, explaining the broadening of the faint end turn over towards fainter UV luminosities with decreasing redshift.

Adding radiative feedback, we find the star formation in low-mass galaxies, that are located in ionized regions, to be increasingly suppressed as reionization proceeds. Hence, the faint end of the UV LF becomes flatter and pushes the turn over at the faint end to lower UV luminosities from $z=10$ to $z=5$ (see {\it Photoionization} and all {\it Heating} models). These trends become stronger towards lower redshifts due to two reasons: firstly, as reionization proceeds (and more regions become ionized), a larger fraction of low-mass halos is affected by radiative feedback. Secondly, the effect of radiative feedback increases as the time when the region became ionized lies further in the past (except the {\it Jeans Mass} model); the ionized and heated gas in the galaxy has had more time to adjust to its lower ``equilibrium'' density.
The higher is the temperature that the IGM is heated up upon ionization (i.e. the higher the ratio between filtering and Jeans mass at virial over-density), the lower is this ``equilibrium'' density and hence the stronger is the effect of the radiative feedback. Indeed, in Fig. \ref{fig_UV_LFs} we see that the turn over at the faint end moves to fainter UV luminosities and that its slope flattens as radiative feedback becomes stronger, i.e. going from the {\it Minimum} and {\it Weak Heating} models to the {\it Photoionization} model\footnote{We note that the shift of the turn over and the flattening of the slope in the {\it Weak Heating} and {\it Photoionization} models are at UV luminosities lower than those that are obtained from converged halos ($M_h\geq10^{8.6}\msun$) in the {\sc vsmdpl} simulation. However, these trends persist even when our model is run on N-body simulations with a $20\times$ higher mass resolutions, such as the {\sc esmdpl} simulation (see Appendix \ref{app_resolution}).} to the {\it Strong Heating} model (and {\it Jeans Mass} model, however we caution the reader that the {\it Jeans Mass} model assumes an instantaneous drop in gas density and is discussed in the following). 

From Fig. \ref{fig_UV_LFs}, we also find that a stronger feedback goes in hand with a higher UV luminosity below which radiative feedback suppresses star formation (indicated by the UV luminosity below which the UV LF starts diverging from the {\it Minimum} model that shows no effects of radiative feedback at $z\gtrsim6$); in the following we refer to this characteristic suppressed UV luminosity as $M_\mathrm{UV,s}$. At each redshift, the value of $M_\mathrm{UV,s}$ corresponds to the UV luminosity of a halo whose mass equals the characteristic mass of the respective radiative feedback model for assuming a reionization redshift of $z_\mathrm{reion}\simeq15$, i.e. when the first progenitors of these $M_\mathrm{UV,s}$ galaxies reionized their environment.
With decreasing redshift the radiative feedback characteristic mass increases and correspondingly $M_\mathrm{UV,s}$ shifts to brighter UV luminosities. 
For example, $M_\mathrm{UV,s}$ shifts from $-17.5$ at $z=7$ to $-18$ at $z=5$ for the {\it Strong Heating} model.
In the {\it Heating} models with $M_c=k\times M_F$, the characteristic mass approaches $k$-times the Jeans mass $M_J(z)$ at $1+z<(1+z_\mathrm{reion})/3.2$ where $z_\mathrm{reion}$ is the reionization redshift. This convergence towards the Jeans mass at virial over-density is also noticeable when comparing the {\it Strong Heating} model with an effective temperature of $T_0=16\times10^4$~K to the {\it Jeans Mass} model with $T_0=4\times10^4$~K. While $M_\mathrm{UV,s}$ for the {\it Jeans Mass} model corresponds to the UV luminosity of a halo with Jeans mass at all times, $M_\mathrm{UV,s}$ for the {\it Strong Heating} model approaches the Jeans mass at virial over-density only at later times when the gas density has had enough time to adjust to the change in temperature upon reionization.

\subsection{Electron scattering optical depth and neutral fraction history}
\label{subsec_reionization}
\begin{figure}
 \centering
 \includegraphics[width=0.49\textwidth]{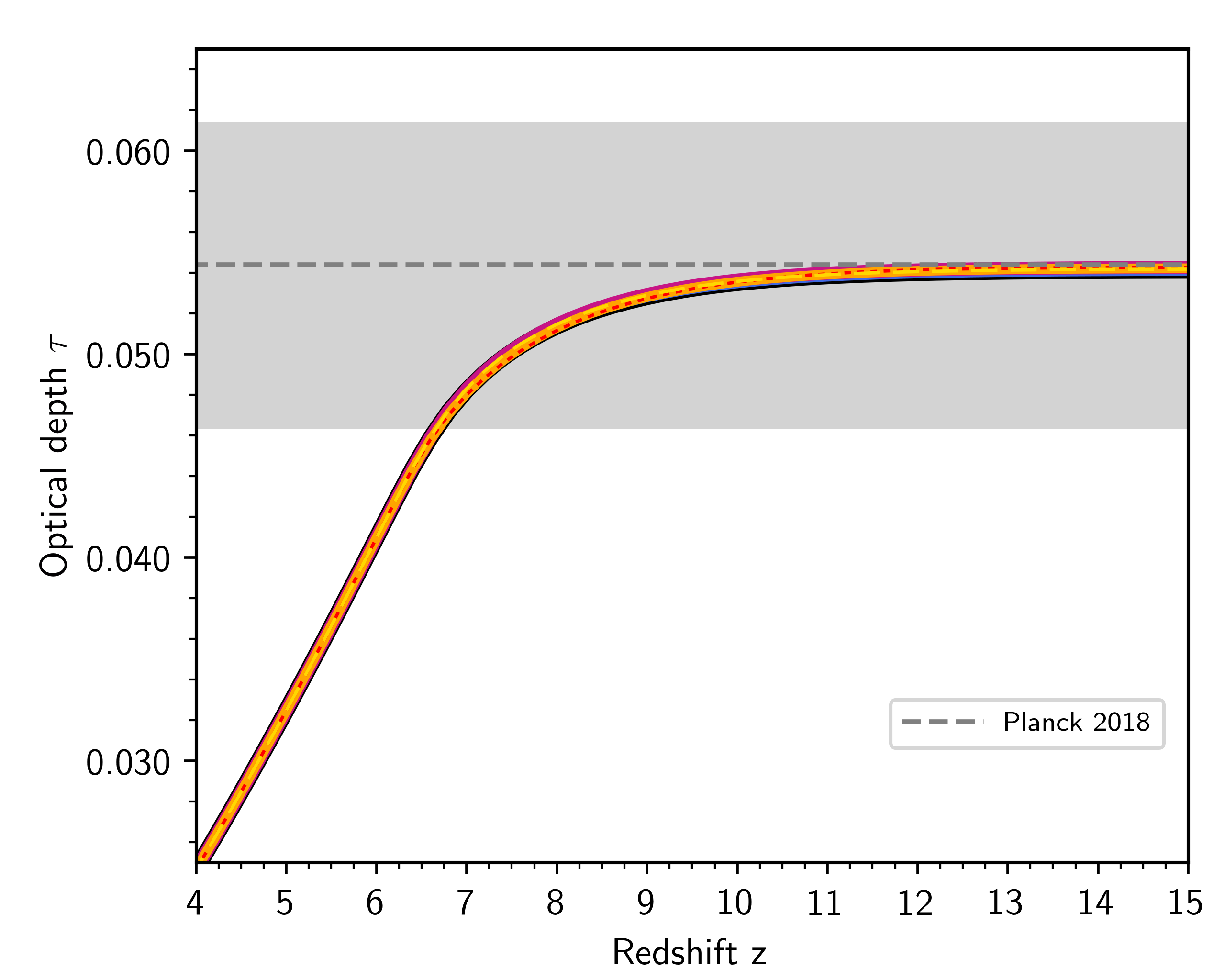}
 \caption{The integrated electron scattering Thomson optical depth using the best-fit parameters noted in Table \ref{tab_best_fit_values} for the different radiative feedback models studied in this work: {\it Minimum} (black solid line), {\it Weak Heating} (blue solid line), {\it Photoionization} (violet solid line), {\it Early Heating} (red dotted line), {\it Strong Heating} (orange solid line) and {\it Jeans Mass} (yellow dashed line). The horizontal grey dashed line shows the central value of the optical depth from Planck \citep{Planck2018} with the grey shaded region showing the associated $1-\sigma$ errors.}
 \label{fig_optical_depth}
\end{figure}

\begin{figure}
 \centering
 \includegraphics[width=0.49\textwidth]{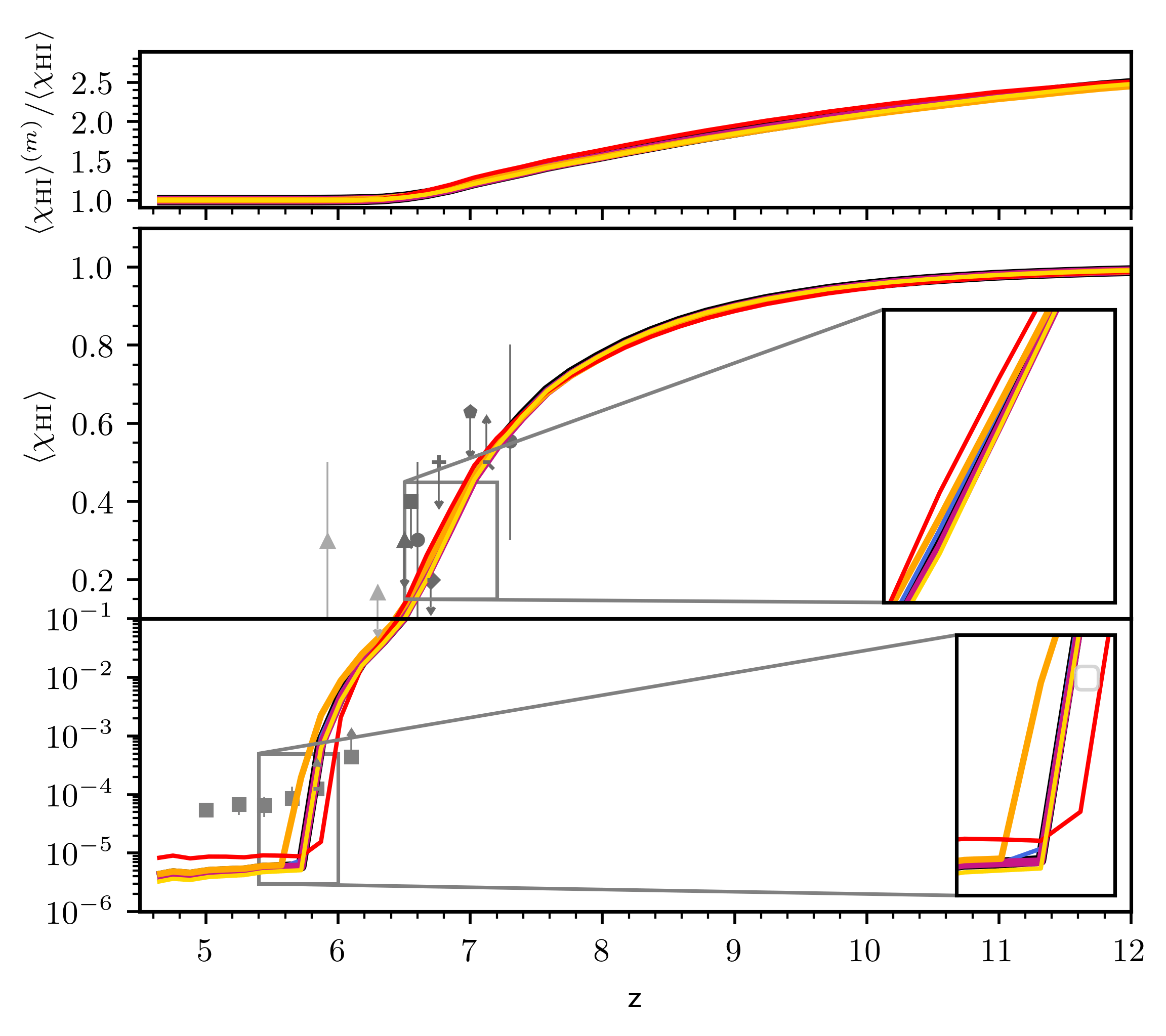}
 \caption{Ratio of the mass- and volume-averaged neutral hydrogen fraction (top panel) and volume averaged neutral hydrogen fraction (bottom panel) as a function of redshift using the best-fit parameters noted in Table \ref{tab_best_fit_values}. In each panel we show results for the different radiative feedback models studied in this work: {\it Minimum} (black solid line), {\it Weak Heating} (blue solid line), {\it Photoionization} (violet solid line), {\it Early Heating} (red solid line), {\it Strong Heating} (orange solid line) and {\it Jeans Mass} (yellow solid line). In the lower panel, grey points indicate observational constraints from: GRB optical afterglow spectrum analyses \citep[light triangles;][]{Totani2006, Totani2014}, quasar sightlines \citep[Medium squares;][]{Fan2006}, Lyman-$\alpha$ LFs  \citep[dark circles]{Konno2018}, \citep[dark squares;][]{Kashikawa2011}, \citep[dark diamonds][]{Ouchi2010}, \citep[dark pentagons][]{Ota2010} and \citep[dark triangles][]{Malhotra2004}, Lyman-$\alpha$ emitter clustering \citep[dark plus signs;][]{Ouchi2010} and the Lyman-$\alpha$ emitting galaxy fraction \citep[dark crosses;][]{Pentericci2011, Schenker2012, Ono2012, Treu2012, Caruana2012, Caruana2014, Pentericci2014}. }
 \label{fig_hist_ion}
\end{figure}

From the {\sc cifog} ionization fields and the corresponding density fields, we derive the global volume-averaged ($\langle\chi_\mathrm{HI}\rangle$) and mass-averaged ($\langle\chi_\mathrm{HI}\rangle^\mathrm{(m)}$) reionization histories as
\begin{eqnarray}
 \langle\chi_\mathrm{HI}\rangle &=& \frac{1}{N_\mathrm{cell}} \sum_{i=1}^{N_\mathrm{cell}} \chi_\mathrm{HI,i} \\
 \langle\chi_\mathrm{HI}\rangle^\mathrm{(m)} &=& \frac{1}{N_\mathrm{cell}} \sum_{i=1}^{N_\mathrm{cell}} \chi_\mathrm{HI,i} \frac{\rho_i}{\langle\rho\rangle}.
\end{eqnarray}
Here $\chi_\mathrm{HI,i}$ and $\rho_i$ are the neutral hydrogen fraction and density in cell $i$, while $N_\mathrm{cell}$ and $\langle\rho\rangle$ are the number of grid cells and the mean density of our simulation box.
Further, the Thomson integrated electron scattering optical depth is calculated as
\begin{eqnarray}
 \tau(z) &=& \sigma_T\ \int_0^z \mathrm{d}z'\ n_e(z')\ \frac{c}{(1+z') H(z')},
\end{eqnarray}
with $\sigma_T = 6.65\times10^{-25}$~cm$^{-2}$ and $H(z)$ being the Thomson cross section and the Hubble parameter at redshift $z$, respectively. The electron number density $n_e$ is determined by the mass-averaged ionization fraction $\langle\chi_\mathrm{HI}\rangle^{(m)}(z)$ and the hydrogen and helium number densities, $n_\mathrm{H}(z)$ and $n_\mathrm{He}(z)$. We assume that the fraction of singly ionized helium equals the fraction of ionized hydrogen, and that helium is fully ionized at $z<3$ \citep[see e.g.][and references therein]{Kulkarni2019}.

As noted before, for each radiative feedback model the $f_\mathrm{esc}$ value (see Table. \ref{tab_best_fit_values}) has been adjusted to reproduce the optical depth $\tau(z_\mathrm{dec})$ for reionization (see Fig. \ref{fig_optical_depth}). The resulting reionization histories are in agreement with existing constraints from quasars, Lyman-$\alpha$ emitters and GRBs (see Fig. \ref{fig_hist_ion}).

In the beginning of reionization, radiative feedback has nearly no impact on the number of ionizing photons emitted and hence the evolution of $\langle\chi_\mathrm{HI}\rangle$ (except for the {\it Jeans Mass} model). However, as reionization proceeds, radiative feedback increasingly suppresses star formation in a rising number of low-mass galaxies as more and more ionized regions form. For a fixed optical depth (c.f. Fig. \ref{fig_optical_depth}), we see from Fig. \ref{fig_hist_ion} that the reionization history becomes more extended, as the strength of the radiative feedback increases (from the {\it Minimum}, {\it Weak Heating} and {\it Photoionization} to the {\it Strong Heating} models). This is because a stronger radiative feedback model causes suppression of star formation in  increasingly massive galaxies, leading to a larger reduction in the number of ionizing photons emitted. As a result, reionization slows down and the Universe becomes fully reionized later (c.f. {\it Photoionization} and {\it Strong Heating} models in Fig. \ref{fig_hist_ion}).

Given this trend, it may seem surprising that the {\it Jeans mass} model, representing one of our strongest radiative feedback models, shows the same redshift evolution of $\langle\chi_\mathrm{HI}\rangle$ as the {\it Minimum} or {\it Weak Feedback} model. The reason for this behaviour originates from the fact that the radiative feedback strength does not increase relative to the SN feedback strength over time, i.e. the ratio of radiative and SN feedback characteristic masses remains constant (see Fig. \ref{fig_filtering_masses}). Hence, the ratio between the ionizing emissivity and halo mass is only lowered by a constant factor, and the lower production of ionized photons in low-mass halos can be compensated by an overall higher escape fraction $f_\mathrm{esc}$.

Moving from a constant $f_\mathrm{esc}$ scenario to one where $f_\mathrm{esc}$ decreases with halo mass, reionization is driven more by galaxies in low-mass halos, resulting in ionized regions being more centred around the smallest over-density peaks in the density field (c.f. higher mass-to-volume averaged neutral hydrogen fraction in top panel of Fig. \ref{fig_hist_ion}). 
Since our simulations are tuned to reproduce the same optical depth value which depends on the mass averaged ionization fraction (and thus on the correlation strength between the IGM density and the time of reionization $z_\mathrm{reion}$), we find the model where $f_\mathrm{esc}$ scales with the ejected gas fraction to accelerate towards the end of reionization, resulting in an earlier ionization of the IGM. 

We briefly note that although the reionization histories shown account for galaxies with halo masses below our convergence limit of $M_h=10^{8.6}\msun$, we do not find the reionization histories to differ noticeably when running our {\sc astraeus} model on an N-body simulation with a $\sim20\times$ better mass resolution (and a convergence limit of $M_h=10^{7.4}\msun$ for 50 DM particles). 

\section{The impact of radiative feedback on early galaxy populations}
\label{sec_SFR}
\begin{figure}
 \centering
 \includegraphics[width=0.49\textwidth]{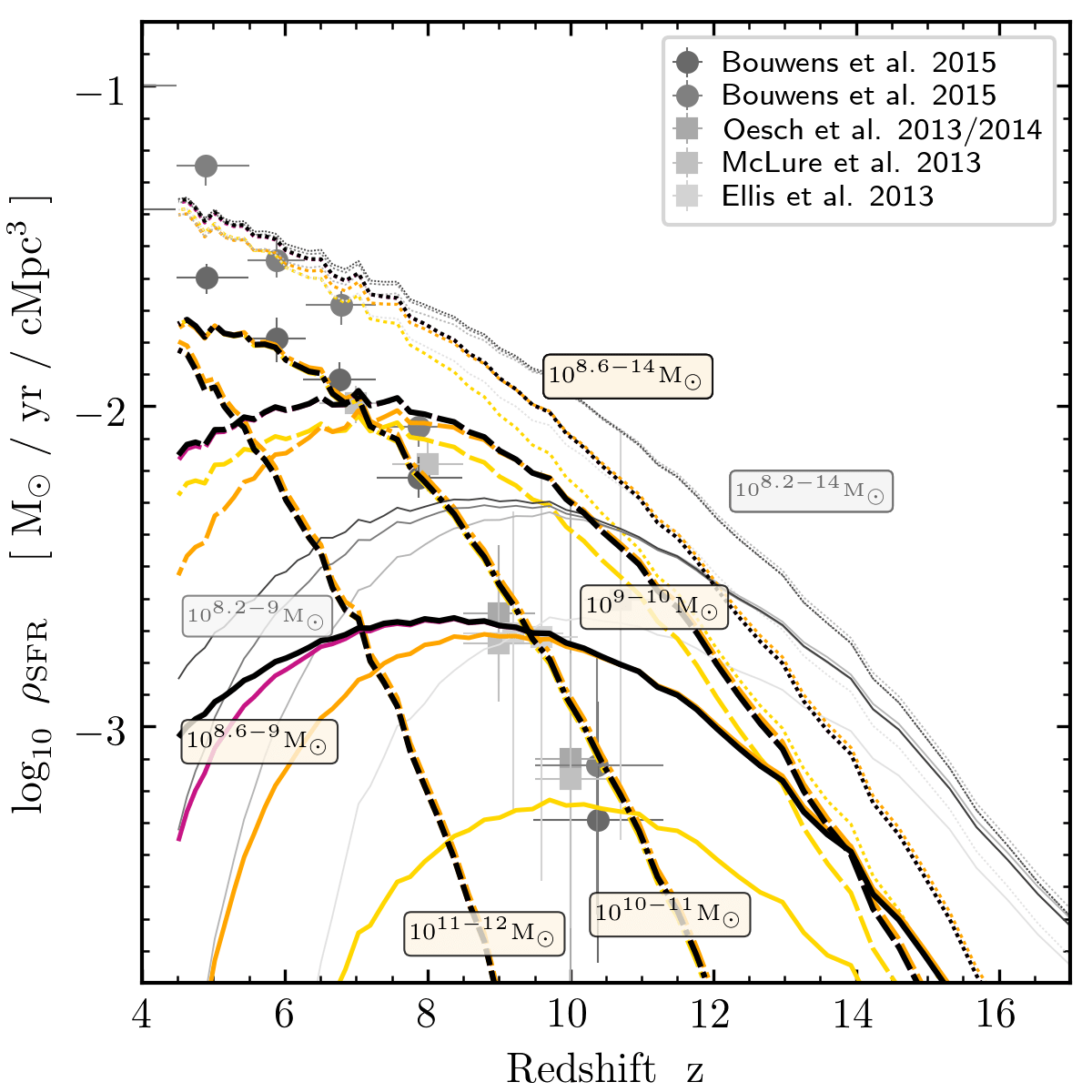}
 \caption{Star formation rate density (SFRD) as a function of redshift for the different halo mass bins marked. We show results for the different radiative feedback models studied in this work: {\it Minimum} (black lines, dark grey lines), {\it Photoionization} (violet lines, medium dark grey lines), {\it Strong Heating} (orange lines, medium bright grey lines) and {\it Jeans Mass} (yellow lines, bright grey). Dotted, solid, long-dashed, dash-dotted, short dashed lines indicate the SFRD for $M_h=10^{8.6-14}\msun$, $10^{8.6-9}\msun$, $10^{9-10}\msun$, $10^{10-11}\msun$, $10^{11-12}\msun$, respectively. Grey lines show the SFRD for $M_h=10^{8.2-14}\msun$ and $M_h=10^{8.2-9}\msun$, including galaxies where the SFR and stellar mass have not fully converged. Grey points show the observational data collected for $M_\mathrm{UV}\leq-17$ LBGs from \citet{Bouwens2015}, \citet{Oesch2013}, \citet{Oesch2014}, \citet{McLure2013} and \citet{ Ellis2013}, as marked.} 
 \label{fig_SFRD_binnedMvir}
\end{figure}

\begin{figure*}
 \centering
 \includegraphics[width=0.9\textwidth]{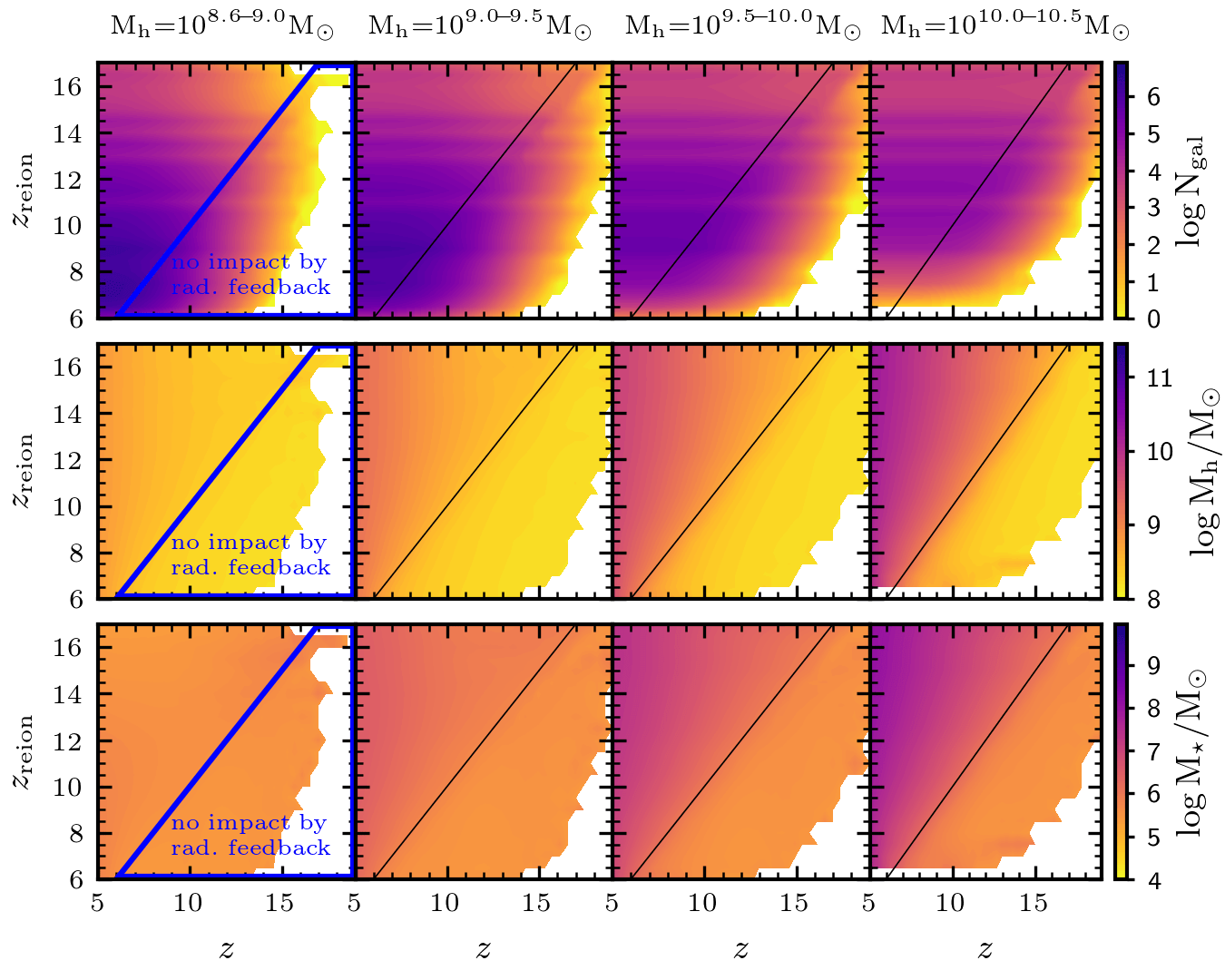}
 \caption{As a function of redshift, we show the number of galaxies (first row), the averaged halo mass histories (second row), and averaged stellar mass histories (third row) binned by the reionization redshift of the galaxy for the {\it Photoionization} model. We show results for different final halo masses at $z=5$, as marked above each column. For a given halo mass $M_h$ and reionization redshift $z_\mathrm{reion}$ bin we average over the halo or stellar mass summed over all its progenitors at redshift $z$. The black solid line marks $z=z_\mathrm{reion}$.}
 \label{fig_SFR_radFb_SOBACCHI}
\end{figure*}

\begin{figure*}
 \centering
 \includegraphics[width=0.9\textwidth]{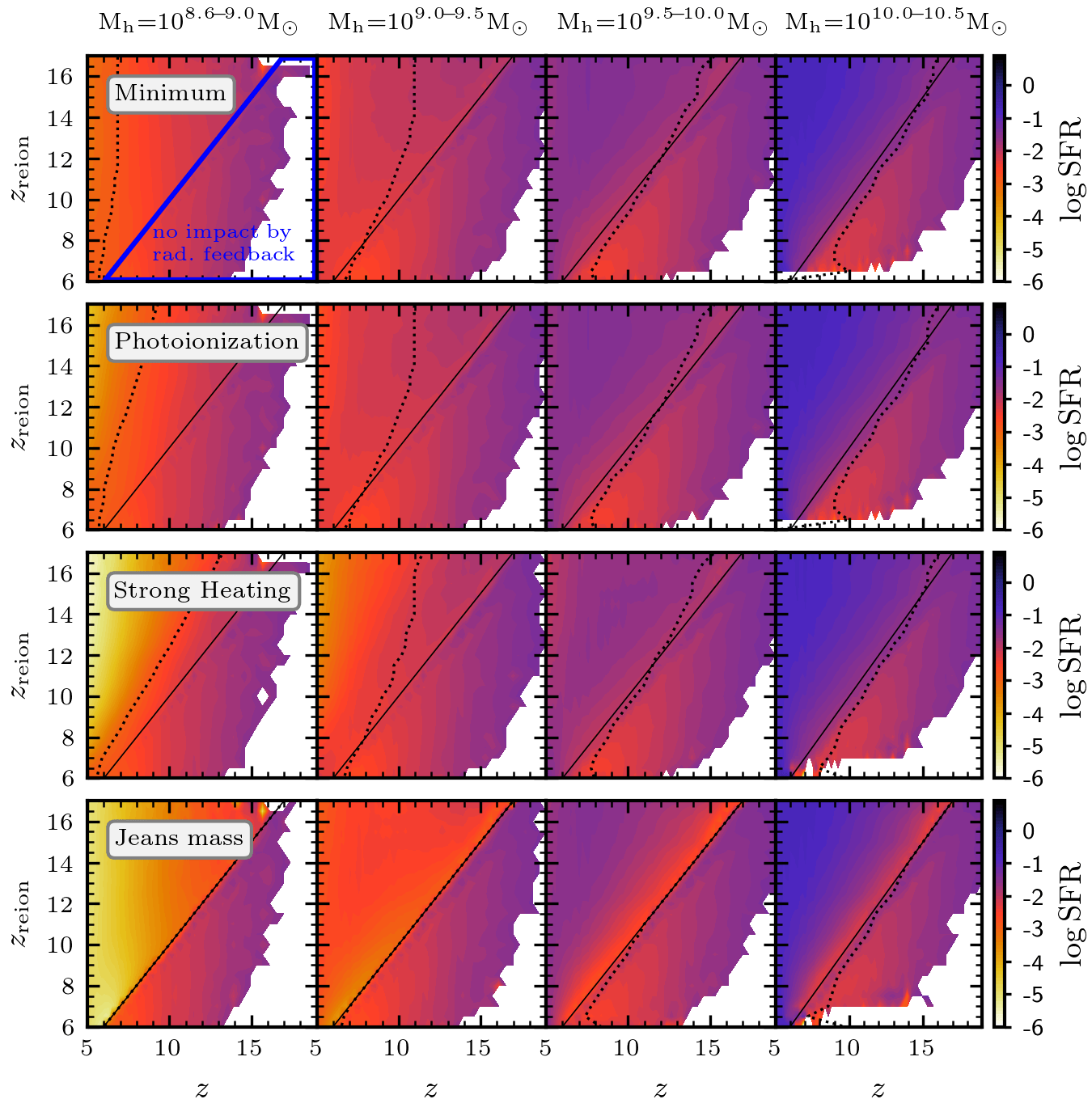}
 \caption{Average star formation rate histories for our different radiative feedback models. Columns show the SFHs of galaxies with different final halo masses $M_h$ at $z=5$. In each panel, halos with final halo mass $M_h$ have been binned according to their redshift at which the galaxy became reionized. For a given halo mass $M_h$ and reionization redshift $z_\mathrm{reion}$ bin we average over the SFR summed over all its progenitors at redshift $z$. The black solid line marks the point where $z=z_\mathrm{reion}$. The black dotted line shows the maximum redshift $z$ where either the average halo mass reaches a mass of $M_h=10^{8.6}\msun$ or the upper limit of the respective halo mass bin exceeds the radiative feedback characteristic mass $M_c(z)$ in the models with radiative feedback.}
 \label{fig_SFR_radFb}
\end{figure*}

In Section \ref{subsec_Mc} we have seen that the degree by which star formation in a galaxy is suppressed by radiative feedback depends sensitively on its individual reionization history and the redshift evolution of the characteristic mass for radiative feedback ($M_c$). Here we explore how the star formation histories of a representative sample of galaxies depend on their location in the cosmic web, as a function of their gravitational potentials and reionization histories.

\subsection{Global star formation rate density for different galaxy masses}
\label{subsec_global_SFR}

We start by discussing the global star formation rate density (SFRD) for galaxies with different halo masses, shown in Fig. \ref{fig_SFRD_binnedMvir}. 
As expected from hierarchical galaxy formation, low-mass halos appear earlier and are always more abundant than high-mass halos throughout cosmic time. While at high redshifts ($z\gtrsim12-13$) low-mass halos (with $M_h=10^{8.2-9}\Msun$) provide the majority (about $60$\% at $z\gtrsim12-13$) of the star formation rate density in all our models \footnote{While the SFR of the lowest mass galaxies in the {\sc vsmdpl} simulation is slightly overestimated due to their initial starburst upon formation, we also recover this trend in the {\sc esmdpl} simulation which resolves $20$ particle halos down to $M_h=10^{7}\msun$.}, more massive halos ($M_h\gtrsim10^9\msun$) start to dominate the SFRD as time proceeds ($z\lesssim11-12$). The reason for this discrepancy between halo abundance and SFRD are radiative and SN feedback processes that contribute to the suppression of star formation in low-mass halos.

Indeed, from Fig. \ref{fig_SFRD_binnedMvir} we can see that for the lowest-mass galaxies ($M_h=10^{8.2-9}\Msun$, solid lines), the SFRD rises with time at $z\gtrsim10$ before turning over at $z\simeq9-10$. The initial rise of the SFRD is due to the increasing number of galaxies that emerge with time. However, as time proceeds, SN feedback suppresses star formation in increasingly more massive galaxies: the location of the peak in the SFRD, $z\simeq9-10$, marks the time when at least half of these low-mass galaxies are fully affected by SN feedback, i.e. their entire gas mass is ejected (see Fig. \ref{fig_filtering_masses}) in addition to these low-mass galaxies inheriting a small, or even no, gas content from their progenitors.
We briefly note that although the halo masses of these galaxies  are close to the mass resolution limit of the underlying {\sc vsmdpl} N-body simulation (and hence their SFR are not fully converged due to the limited depth of their merger trees), we find the position of the turn over to prevail when these low-mass galaxies are better resolved.
While the drop in the SFRD at $z\lesssim9$ is entirely due to SN feedback for the {\it Minimum} model, models including radiative feedback show an additional drop in the SFRD. In case of the {\it Heating} and {\it Photoionization} models this drop increases with time, since radiative feedback causes an additional suppression of star formation in a rising number of galaxies as an increasing fraction of the IGM becomes ionized. In contrast, the {\it Jeans Mass} model shows an overall lower SFRD with an increasing logarithmic difference with decreasing redshift compared to the other models. This increasing logarithmic difference in the SFRD with decreasing redshift is caused by the increasing number of low-mass galaxies (below the Jeans mass) in ionized regions that are affected by radiative feedback as soon as their environment is reionized. We emphasize that, although the merger trees of the majority of our lowest-mass galaxies suffer from short lengths, the drop in the SFRD towards lower redshifts in our different radiative feedback models persists even when the lengths of the merger trees are longer and the underlying N-body simulation has a better mass resolution.

For more massive galaxies, $M_h=10^{9-10}\msun$ (long dashed lines) and $M_h=10^{10-11}\Msun$ (dash-dotted lines), the turn over in the SFRD at lower redshifts, $z\simeq7-8$ and $(4-5)$, is also caused by the increasing impact of SN feedback in decreasing the gas mass brought in by their merging progenitors with cosmic time. However, in contrast to the lowest-mass halos, only a fraction of the gas in the galaxies is ejected, shifting the turn-over to increasingly lower redshifts with increasing halo mass. Furthermore, for $M_h=10^{9-10}\Msun$, we see that only the strongest radiative feedback models ({\it Strong Heating} and {\it Jeans mass}) cause a noticeable decrease in the SFRD. Only in those models, the radiative feedback characteristic mass exceeds $10^9\Msun$ significantly for galaxies whose environment was reionized at $z\simeq7-8$ (when the majority of the volume becomes ionized). While theoretically, the star formation in $M_h=10^{10-11}\Msun$ halos reionized very early on could also be suppressed by radiative feedback in the {\it Strong Heating} model, this is not the case in our reionization scenarios where the IGM was predominantly reionized at $z\lesssim9$ (see also the reionization history in Fig. \ref{fig_hist_ion}).
For high-mass galaxies, the {\it Jeans mass} model shows an overall lower SFRD, albeit with a constant logarithmic offset. While the lower SFRD is again due to the immediate effect of radiative feedback upon reionization, the constant logarithmic offset (and not an increasing one as for low-mass sources) is caused by the fact that the fraction of $10^{9-10}\Msun$ halos affected by radiative feedback remains constant over time. Again, here the lower gas content in $10^{9-10}\Msun$ halos is caused by their progenitors being gas-poorer for stronger radiative feedback.

We note that similar trends of the global SFRDs for various halo mass bins have also been found in the CoDaI simulation \citep[cf.][and see Section \ref{subsubapp_comparison_SFR}]{ocvirk2016, Dawoodbhoy2018}. 

\subsection{Impact of the reionization redshift on the SFR}
\label{subsec_zreion_SFR}

In this section, we start by discussing the (average) redshift-dependent stellar mass ($M_\star$) and halo mass ($M_h$) assembly histories for galaxies at $z=5$ as a function of their reionization redshift, as shown in Fig. \ref{fig_SFR_radFb_SOBACCHI}. In the same figure, we also show the number of galaxies ($N_\mathrm{gal}$) occupying this $z-z_{reion}$ plane, over which the stellar mass and halo mass assembly histories have been averaged. 

We find $N_\mathrm{gal}$ to reflect the hierarchical structure formation scenario, where a massive galaxy at a given time has formed and (in an inside-out reionization scenario) reionized its environment earlier than a less massive galaxy. For $M_h=10^{10-10.5}\msun$ halos at $z=5$, $N_\mathrm{gal}$ remains almost constant at $z<z_\mathrm{reion}$ (see fourth panel in the first row). In contrast, as we go to the least massive galaxies in our simulation, $N_\mathrm{gal}$ starts to rise towards smaller $z$ values on the x-axis (e.g. for $M_h=10^{8.6-9}\msun$ and $z_\mathrm{reion}=12$, there are about $N_\mathrm{gal}\simeq 10^4$ galaxies at $z=11$, while $N_\mathrm{gal}\simeq 10^{5}$ at $z=5$). Whether $N_\mathrm{gal}$ remains constant at $z<z_\mathrm{reion}$ also provides an indication of whether the galaxies in the respective halo mass bin have been able to ionize their grid cell in the simulation alone over the course of their life. From the figure we see that in our simulations the threshold for having emitted enough ionized photons to ionize the respective grid cell lies around $M_h\simeq10^9\msun$, i.e. when star formation is not severely suppressed by feedback.
From the same figure, we also see the build-up of stellar and the halo mass of galaxies. In our simulations a galaxy with $M_h\simeq10^{10.5}\msun$ (with $M_\star\simeq10^8\msun$) at $z=5$ forms early on at $z\gtrsim17$ (i.e. $z_\mathrm{reion}\simeq17$) with a mass of $M_h\simeq10^8\msun$ ($M_\star\simeq10^{5.5}\msun$) and continuously accumulates mass reaching $M_h\simeq10^{9.5}\msun$ ($M_\star\simeq10^{7}\msun$) at $z\simeq11$. Furthermore, we note that in our model, reionization proceeds in an inside-out fashion, i.e. the ionization fronts move from over- to under-dense regions. Consequently, $z_\mathrm{reion}$ is correlated to the underlying density field and thus the halo mass. Halo masses close to the upper limit of the halo mass bin are more likely to be found at high $z_\mathrm{reion}$ values, while those close to the lower limit of the halo mass bin have lower $z_\mathrm{reion}$ values. This effect can be seen in the second row in Fig. \ref{fig_SFR_radFb_SOBACCHI} where we show the average halo mass history of $z=5$ galaxies in the given $z_\mathrm{reion}$ and halo mass bins, e.g. for $M_h=10^{10-10.5}\msun$ the final halo mass at $z=5$ is $\sim10^{10}\msun$ for $z_\mathrm{reion}=6$ and $\sim10^{10.5}\msun$ for $z_\mathrm{reion}=15$. In the following we will refer to this effect as the positive $z_\mathrm{reion}-M_h$ correlation effect.
Also, in more massive halos with $M_h\gtrsim10^9\msun$, the stellar mass follows the growth of the halo mass. However, in less massive halos ($M_h\lesssim10^9\msun$), the stellar mass does not follow the growth of the halo mass but remains constant (or even drops for $M_h\lesssim10^{8.6}\msun$) as star formation is increasingly suppressed by SN and radiative feedback (c.f. first panels in Fig. \ref{fig_SFR_radFb_SOBACCHI}).

We now discuss the (average) redshift evolution of the star formation rate histories (SFH) for galaxies in a given halo mass bin at $z=5$ as a function of the redshift $z_\mathrm{reion}$ when the surrounding region of the galaxy became ionized in Fig. \ref{fig_SFR_radFb}. We remark that for $z>z_\mathrm{reion}$ the shown star formation rate (SFR) is averaged over increasingly fewer galaxies with increasing redshift (see last row in Fig. \ref{fig_SFR_radFb_SOBACCHI}) and not affected by radiative feedback. Given our interest in studying the impact of radiative feedback on the SFH, we limit our discussion to $z\leq z_\mathrm{reion}$ (i.e. galaxies above the black solid line). 
Furthermore, we limit our discussion of the SFHs to the area left of the black dotted line in each panel of Fig. \ref{fig_SFR_radFb}. This dotted line indicates the maximum redshift $z$ where either the average halo mass reaches a mass of $M_h=10^{8.6}\msun$\footnote{We note that at this limit roughly half of the SFHs have converged.} or the upper limit of the respective halo mass bin exceeds the radiative feedback characteristic mass $M_c(z)$. The first condition indicates where the analysed average SFHs are converged. However, this convergence can also be reached at lower average halo mass (or higher redshifts $z$), if the halo masses of galaxies exceed the radiative feedback characteristic mass (second condition). The SFRs of the respective galaxies become then strongly constrained by their redshift $z$ and the redshift of reionization $z_\mathrm{reion}$.

Since the {\it Minimum} model effectively corresponds to the case of SN feedback only, the SFR at redshift $z$ is basically independent of $z_\mathrm{reion}$ (when accounting for the $z_\mathrm{reion}-M_h$ correlation effect) and only depends on the mass of the galaxy and its redshift $z$ (first row in Fig. \ref{fig_SFR_radFb}). In agreement with the SN filtering mass (see dash-dotted grey line in Fig. \ref{fig_filtering_masses}), we can see that SN feedback suppresses star formation only in halos with $M_h\lesssim10^{9.5-10}\Msun$ with the suppression for a given $M_h$ increasing with decreasing redshift. For more massive halos at $z=5$, the SN energy is not large enough to push out most of the gas, and the SFR continues to rise with decreasing redshift as galaxies becomes more massive. For galaxies in halos with $M_h=10^{9.5-10}$, we see that the SFR peaks again around $z\simeq7-8$\footnote{We note that if the SFHs are better resolved (as e.g. in the {\sc esmdpl} simulation described in  Appendix \ref{app_resolution}), the SFR peaks also around $z\simeq7-8$ for galaxies in halos with $M_h=10^{9-9.5}$ in agreement with Fig. \ref{fig_SFRD_binnedMvir}.} (see also Fig. \ref{fig_SFRD_binnedMvir}): due to the increasing characteristic mass for SN feedback with decreasing redshift, these halos become increasingly SFR suppressed by SN feedback with time.

Including radiative feedback changes the SFHs, since they become dependent on the reionization redshift $z_\mathrm{reion}$. For low-mass galaxies, the SFR at redshift $z$ decreases as the galaxy is located in a region that has been reionized earlier (higher $z_\mathrm{reion}$ value) due to the higher radiative feedback characteristic mass $M_c$ (see discussion in Section \ref{subsec_Mc}). For example, in the case of the {\it Photoionization} model, $M_h=10^{8.6-9}\msun$ halos at $z=7$ have a SFR $\sim10^{-2.5}\msun ~{\rm yr^{-1}}$ for $z_\mathrm{reion}=7$, while the SFR drops to a value as low as $\sim10^{-3.5}\msun ~{\rm yr^{-1}}$ for $z_\mathrm{reion}=15$.

Furthermore, the strength and time of radiative feedback depends strongly on the gravitational potential of the galaxy. The less massive a galaxy is, the shallower is its potential, the earlier and the more its star formation is suppressed by radiative feedback: from low to high halo masses (left to right in Fig. \ref{fig_SFR_radFb}) the relative drop of the SFR due to radiative feedback (not SN feedback) since its maximum decreases: while the SFR drops by roughly 2-3 (1-2) orders of magnitude for $M_h=10^{8.6-9.5}\Msun$ ($M_h=10^{8.6-9}\Msun$) halos within the area of convergence (left of the black dotted line), there is nearly no drop for $M_h=10^{9.5-10}\Msun$ ($M_h=10^{9-9.5}\Msun$) in the {\it Strong Heating} ({\it Photoionization}) model. As the radiative feedback model becomes more efficient, the drop in the SFR becomes not only more pronounced but extends also to galaxies with higher halo masses. While the maximum halo mass being affected by radiative feedback at $z\simeq5$ is $M_h\simeq10^{9}\Msun$ for the {\it Photoionization} model, it increases to $M_h\simeq10^{10}\Msun$ for the {\it Strong Heating} model, respectively. 

\begin{figure*}
 \centering
 \includegraphics[width=0.7\textwidth]{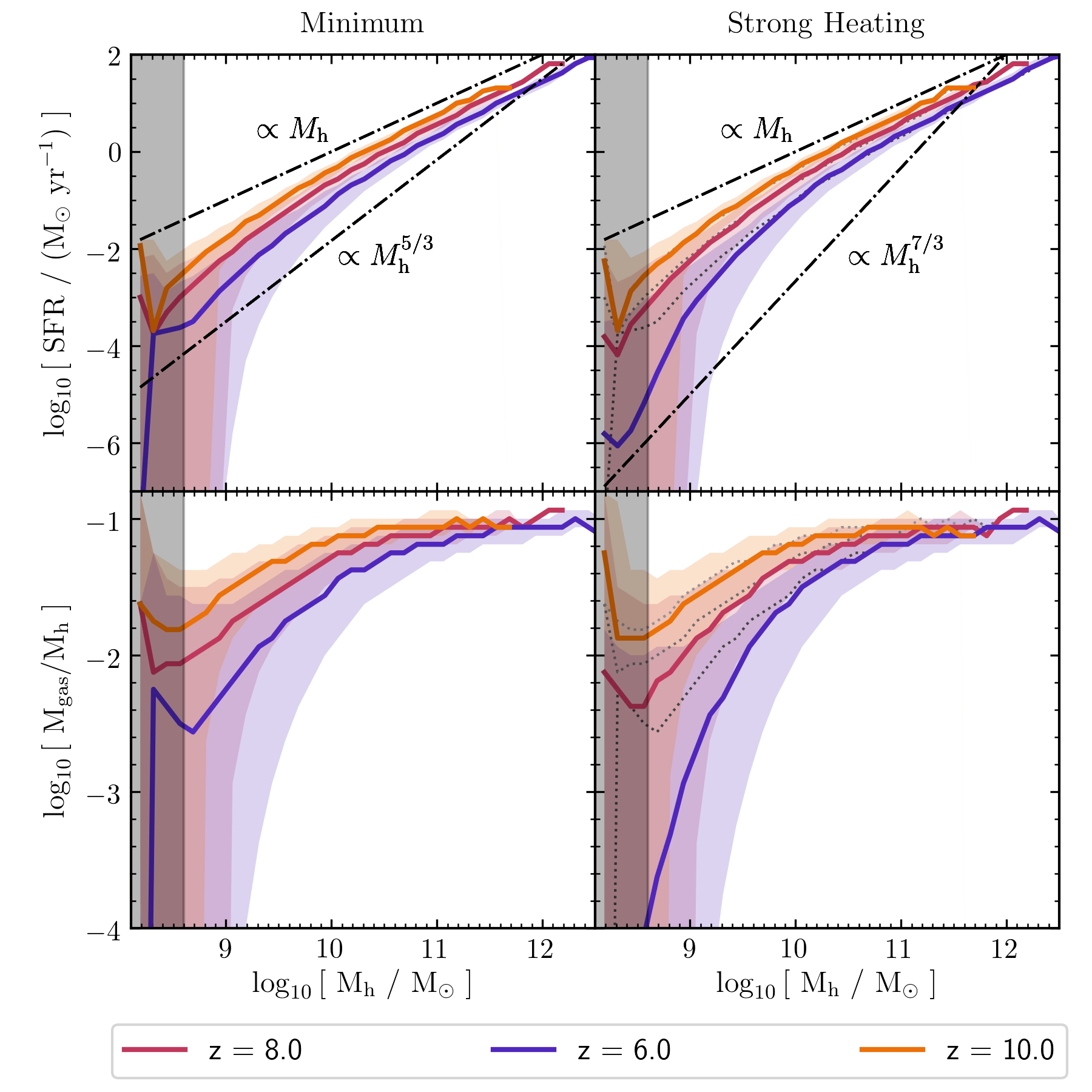}
 \caption{As a function of the halo mass, we show the star formation rate (top row) and the initial gas fraction (bottom row) for the {\it Minimum} model (left column) and the {\it Strong Heating} model (right column). In all panels, we show results at $z=6$ (blue solid line), $z=8$ (red solid line) and $z=10$ (orange solid line). Light grey to black dotted lines show the results for a scenario with SN feedback only ($f_s=0.01$, $f_w=0.2$). The lines represent the median of the distribution and the shaded areas mark the region where 80\% of the galaxies are located. Black dot-dashed lines show the indicated halo mass proportionalities to allow easy comparisons with relations found in \citet[][$\mathrm{SFR}\propto M_h^{7/5}$]{Mutch2016} and \citet[][$\mathrm{SFR}\propto M_h^{5/3}$]{ocvirk2016}. The grey shaded area marks the halo masses that is affected by the mass resolution limit of the underlying N-body simulation.}
 \label{fig_fg}
\end{figure*}

We also note that the SFHs for low-mass ($10^{8.6-9}\Msun$) and mid-mass ($10^{9-10}\Msun$) halos differ from the point of reionization at $z_\mathrm{reion}\gtrsim10$: while the SFR drops continuously for low-mass halos (see $M_h\lesssim10^9\Msun$ for ({\it Photoionization} and) {\it Heating} models), it rises first and drops then for mid-mass halos (see $10^{9.5}\Msun\lesssim M_h\lesssim10^{10}\Msun$ for {\it Heating} models). These trends are also seen when the merger trees and hence the SFHs of low-mass galaxies are better resolved, and are consistent with the findings from the radiation hydrodynamic simulation CoDaI \citep{Dawoodbhoy2018}.
Besides radiative feedback, galaxies in low-mass galaxies are also subject to SN feedback, which causes the SFR to decline from their start. In contrast, galaxies in mid-mass halos are only marginally affected by SN feedback but become with decreasing redshift subject to the time-delayed radiative feedback in the {\it Photoionization} and {\it Heating} models: the SFR keeps rising as long as the halo mass of the galaxy exceeds the radiative feedback characteristic mass $M_c$, and starts dropping continuously as soon as $M_c$ has surpassed the galaxy's halo mass.
Since $M_c$ is a function of the reionization redshift $z_\mathrm{reion}$ of a chosen galaxy, the resulting peak in the SFR also depends on $z_\mathrm{reion}$. As the strength of radiative feedback of a model increases, its redshift shifts closer to $z_\mathrm{reion}$, since $M_c$ surpasses the considered halo mass shorter after the reionization of the galaxy.
During reionization, we find that $M_c$ never approaches the halos with $M_h\gtrsim10^{10}\Msun$, and consequently SFHs are not affected by radiative feedback.

Finally, we comment on the results from the {\it Jeans Mass} model. This model differs from the {\it Photoionization} and {\it Heating} models, since the degree of star formation suppression does not depend on when a galaxy's environment became reionized ($z_\mathrm{reion}$) but on its redshift and the galaxy's halo mass (similar to SN feedback). Consequently, although this model results in a strong suppression of star formation in low mass halos, the SFHs effectively mimic a higher SN characteristic mass.

\subsection{Impact of the reionization redshift on the gas fraction}
\label{subsec_baryon_fraction}

In this Section we discuss how the relation between the halo mass and star formation rate and halo mass and the initial gas-fraction $M_g^i/M_h$ evolve with redshift and the reionization feedback model used. As we have detailed in Sec. \ref{subsec_gal_model}, the halo mass is determined by mergers and accretion, while the initial gas mass (i.e. gas mass present after mergers and accretion but before star formation and SN feedback) and hence the star formation rate depends on the star formation and assembly history of a galaxy.

While in real galaxies, processes such as gas accretion, gas ejection and star formation overlap at times, the complex interplay of the physical processes in galaxies impedes a direct all-encompassing solution but forces us to execute them sequentially. Hence, in our model, gas accretion (including mergers) and evaporation happen at the beginning of a time step, while gas ejection and star formation is executed at the end. In this Section, we comment on the initial gas fraction $M_\mathrm{g}^i/M_h$.

In Fig. \ref{fig_fg} we show the median SFR (top panels) and gas fractions (bottom panels) as a function of the halo mass $M_h$ for the {\it Minimum} and {\it Strong Heating} models (left and right panels, respectively).
We see that the gas fraction and SFR show the same increasing trend with halo mass $M_h$ (c.f. bottom and top panel in Fig. \ref{fig_fg}) as expected from our model (see Section \ref{subsubsec_star_formation}). However, while the SFR continuously rises with increasing $M_h$, the gas fraction increases with $M_h$ before saturating to a value that is around 60\% of the cosmological baryon fraction $\Omega_b/\Omega_m$ at $M_h\simeq10^{10.5-11.5}$ for $z=10-6$. While the SFR is a direct tracer of the gas mass in the galaxy, the increasing gas fraction with rising halo mass reflects the deepening of the gravitational potential and the associated decrease in the efficiency with which SN explosions can eject gas from galaxies. The saturation of the gas fraction to a value lower than the cosmological baryon fraction towards and for massive galaxies stems from the gas loss of their lower mass progenitors.
When radiative feedback is added, we find the gas fraction to decrease, with the decrease being strongest for the lowest halo masses (c.f. {\it Strong Heating} to {\it Minimum} model in Fig. \ref{fig_fg}). Galaxies in shallower potentials have shorter dynamical timescales and hence they are the quickest to adjust their gas density to the heating by reionization and to increase their Jeans mass or radiative feedback characteristic mass. Similarly, this build-up of the galaxy's actual Jeans mass (or increasing radiative feedback characteristic mass) leads to a stronger drop in the gas fractions, relative to the {\it Minimum} model, as redshift decreases.

Finally, we find our gas fractions to be lower than those obtained from radiation hydrodynamical simulations \citep[e.g.][]{Hasegawa2013, Okamoto2008, Gnedin2014, Pawlik2015, Wu2019}, since our semi-analytical galaxy evolution model accounts only for the kinetic but not temperature effects of SN feedback, i.e. all the gas that does not form stars can be ejected out of the potential rather than just being heated as in radiation hydrodynamical simulations. Ignoring gas ejection from SN feedback and accounting only for gas evaporation and reduced accretion from radiative feedback ($f_g \Omega_b/\Omega_m$ versus $M_h$), our model yields results comparable to those of radiation hydrodynamical simulations. A more detailed discussion can be found in the Appendix \ref{subsubapp_comparison_baryon_fraction}. 

\begin{figure*}
 \centering
 \includegraphics[width=0.99\textwidth]{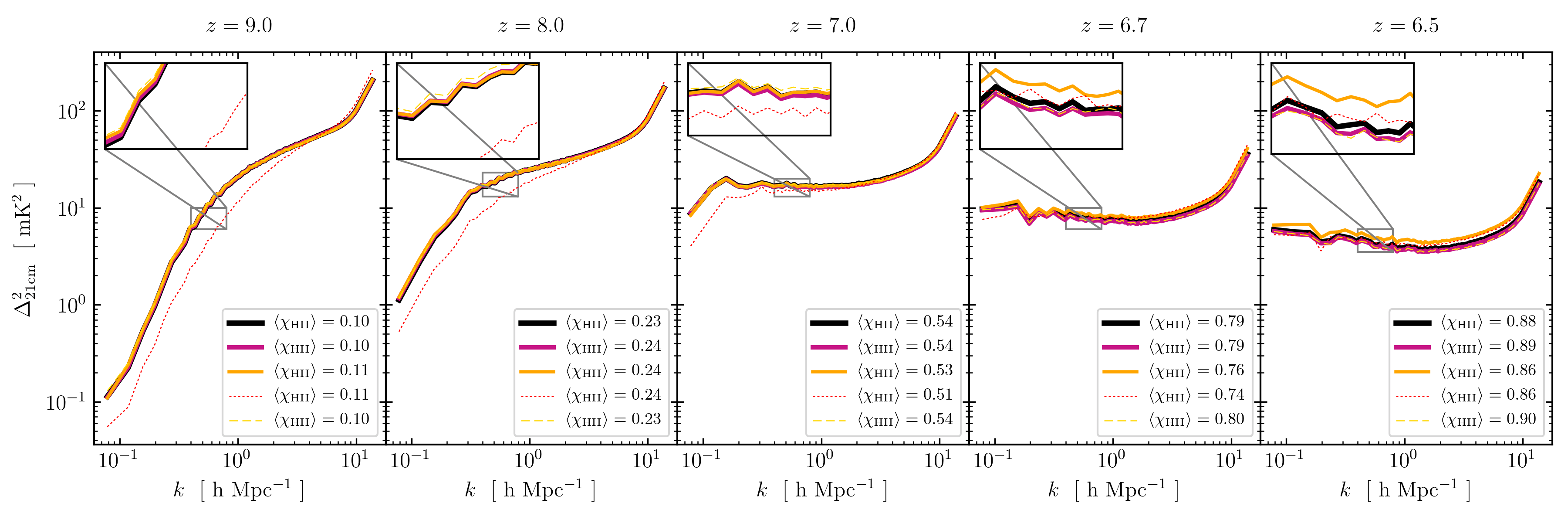}
 \caption{21cm power spectra at fixed redshifts $z$  using the best-fit parameters noted in Table \ref{tab_best_fit_values}. In each panel we show results for the different radiative feedback models studied in this work: {\it Minimum} (black solid line), {\it Photoionization} (violet solid line), {\it Early Heating} (red dotted line), {\it Strong Heating} (orange solid line) and {\it Jeans Mass} (yellow dashed line). In each panel, we also show the average volume-averaged \HI fraction in each model at that redshift.}
 \label{fig_ps21cm_redshift}
\end{figure*}

\begin{figure*}
 \centering
 \includegraphics[width=1.01\textwidth]{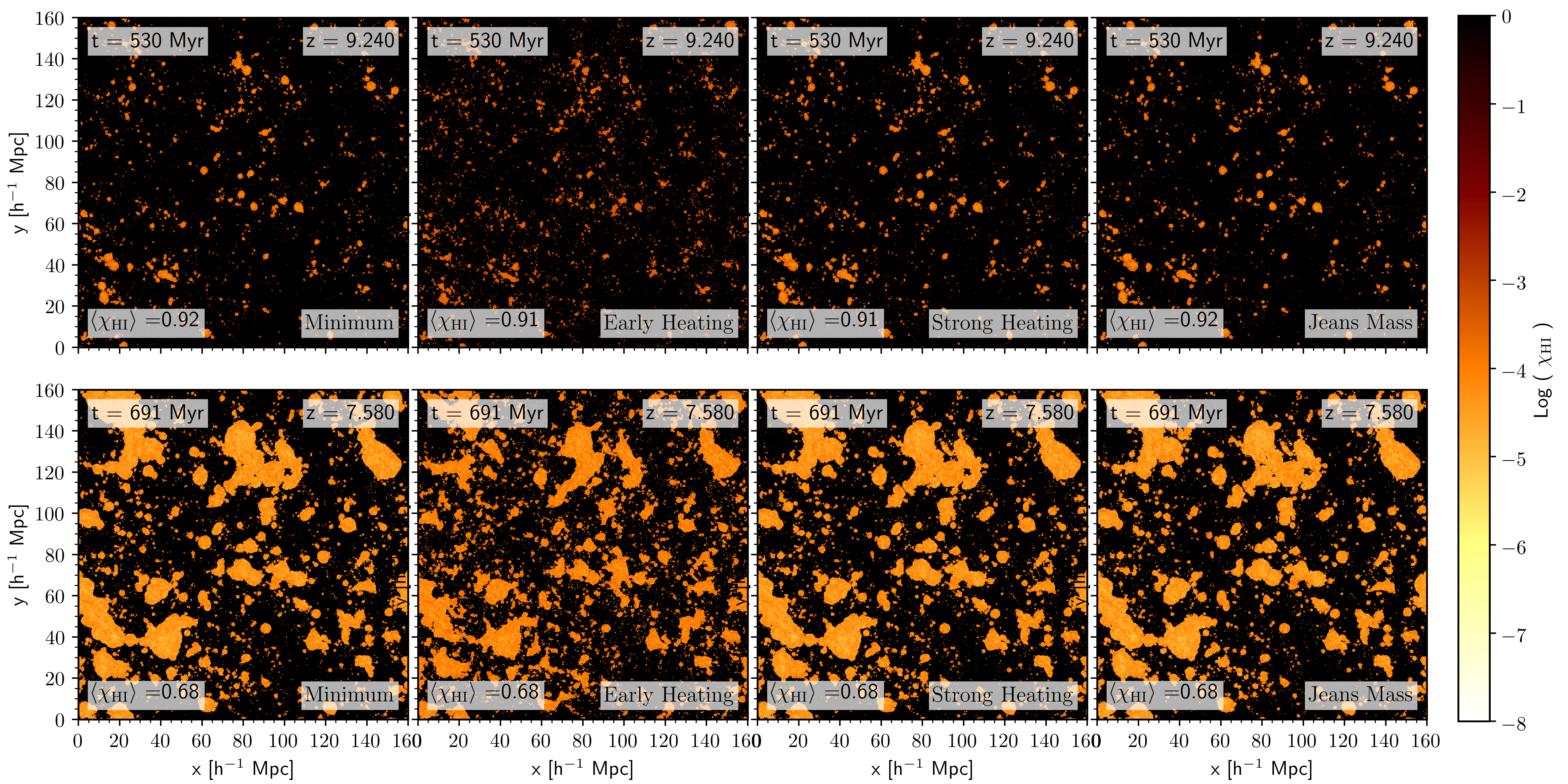}
 \caption{Neutral hydrogen fraction fields at $z=9.2$ (top) and $z=7.6$ (bottom) of the {\it Minimum}, {\it Early Heating}, {\it Strong Heating} and {\it Jeans Mass} models (from left to right). The volume-averaged value of the neutral fraction in each cell is marked in the panels. For each redshift and model, we show a slice through the centre of the simulation box.}
 \label{fig_XHI_fields}
\end{figure*}

\section{The impact of radiative feedback on the 21cm power spectrum}
\label{sec_21cm_ps}

Current and forthcoming radio interferometers will provide measurements of the 21cm signal that arises from the spin flip transition in neutral hydrogen \HI. One of the key quantities to be measured is the 21cm power spectrum that provides a tracer of the size distribution of the ionized and neutral regions (see \citet{Hutter2020} for the derivation of the 21cm signal from the simulations\footnote{We assume that the IGM has been heated by X-rays, i.e. the spin temperature is well heated above the CMB temperature during reionization.}). Here we discuss whether radiative feedback has an impact on the sizes of the ionized regions and the associated redshift evolution of the 21cm power spectra. For this purpose, we show the 21cm power spectra at redshifts $z=9.0$, $8.0$, $7.0$, $6.7$ and $6.5$ (Fig. \ref{fig_ps21cm_redshift}) corresponding roughly to $\langle \chi_\mathrm{HII}\rangle \sim 0.1$, $0.25$, $0.5$, $0.75$ and $0.9$, respectively, as well as the ionization maps of our simulation slices at $z=9.2$ and $z=7.6$ (Fig. \ref{fig_XHI_fields}).

First, we consider all reionization scenarios where a constant ionizing escape fraction has been assumed (i.e. all except {\it Early Heating}). For these scenarios, the ionization fields and 21cm power spectra are very similar, however we note a few small differences. Firstly, while the 21cm power spectra are effectively identical during the first two thirds of reionization, the increasing suppression of star formation (and hence the emission of ionizing photons) in low-mass galaxies extends reionization as the radiative feedback strength is increased from the {\it Minimum} to the {\it Strong Heating} model as discussed in Sec. \ref{subsec_reionization}. With the amplitude of the 21cm power spectrum tracing the mean neutral hydrogen fraction, we find the 21cm power spectrum amplitude to reflect the reionization histories, with the {\it Strong Heating} model having a marginally higher amplitude than the weaker radiative feedback models and the {\it Jeans Mass} model.
Secondly, in terms of the reionization topology, a stronger radiative feedback translates into less ionizing photons being emitted from low-mass galaxies, resulting in smaller ionized regions around these sources. Focusing on the smallest ionized regions in Fig. \ref{fig_XHI_fields}, one can see that ionized regions of cell size are more ionized and abundant in the {\it Minimum} than in the {\it Jeans Mass} model. In case of the {\it Strong Heating} model, a difference to the {\it Minimum} model is barely seen, as radiative feedback has not become effective at these redshifts.

Across our reionization scenarios, we note that the ionization fields and 21cm power spectra of the {\it Early Heating} model differ strongly from all other models during the first half of reionization: due to the higher (lower) abundance of small-sized (large-sized) ionized regions respectively, the large scale 21cm power is reduced. Given that the ionizing escape fraction in this scenario scales with the ejected gas fraction, and hence decreases with halo mass, this change is expected and is in agreement with previous works \citep[e.g.][]{Seiler2019}. 

If radiative feedback was even stronger than our {\it Strong Heating} model, i.e. star formation in low-mass sources was more suppressed and star formation in even higher-mass galaxies became affected, it would have the two effects. Firstly, this would result in more-small sized ionized regions, since low-mass galaxies forming in neutral regions are not affected by radiative feedback (and delayed SN feedback) in the first $\sim20$~Myrs of their life. Secondly, we would find larger-sized ionized regions to be smaller and less connected, since a stronger radiative feedback reduces star formation - and hence the number of ionizing photons produced - not only in more massive galaxies, but also in low-mass galaxies located in the vicinity to the most massive galaxies that contribute to the connectivity of the ionized regions.

In summary, even the strongest radiative feedback hardly affects the galaxies that determine the ionization topology on $>1-2$~cMpc scales, which is in agreement with previous findings  \citep [e.g.][]{Dixon2016}. Different functional forms of the ionizing escape fraction dominate over the intrinsic ionizing emissivity distribution of the underlying galaxy population and can lead to significantly different ionization topologies. We will explore this finding in more detail in future works.

\section{Impact of stellar population synthesis models}
\label{sec_sps}

In all the previous sections we have presented the results using the single stellar population synthesis (SPS) model {\sc starburst99} \citep{leitherer1999}. Over the past decade, it has been increasingly recognised that the effects of stellar multiplicity can not be neglected when modelling the evolution and observable properties of young stellar populations, since about $70\%$ of massive stars are part of binaries \citep{Sana2012} and exchanging mass with their companions affects their structure and evolution. For this reason, we also have run our simulations where the ionizing emissivity of a starburst is derived with the {\sc bpass} SPS model\footnote{We note that the UV luminosity produced with {\sc bpass} hardly differs from that generated with {\sc starburst99}.} \citep{eldridge2017}. While the initial ionizing emissivity at $\lesssim 4$Myrs is similar in both, {\sc starburst99} and {\sc bpass}, models, at later times the ionizing emissivity decreases slower in time in the {\sc bpass} model (see equations \ref{eq_QSP_starburst99} and \ref{eq_QSP_bpass}). Thus, we find the intrinsic ionizing luminosities of all galaxies to be about a factor $10$ higher. However, to fit the Planck optical depth, this increased ionizing emissivity from galaxies is compensated by a lower escape fraction of these ionizing photons, $f_\mathrm{esc}^\mathrm{BPASS}\simeq 0.1 \times f_\mathrm{esc}^\mathrm{S99}$. This re-normalisation causes the effective ionizing emissivities to be about equal, and reionization histories as well as the sizes, shapes and distribution of the ionized regions agree well with each other. This strong agreement for a constant relation between $f_\mathrm{esc}^\mathrm{BPASS}$ and $f_\mathrm{esc}^\mathrm{S99}$ arises because the number of ionizing photons at each time is dominated by the youngest stellar populations, given that the SFR remains constant within the current time step ($10-30$~Myr) and drops strongly in the subsequent time step.
Since we derive the distribution of the ionized regions from the cumulative number of ionizing photons, we do not expect that smaller time steps would result in significantly differing ionization topologies.

It is interesting to note that \citet{Rosdahl2018} found higher ionizing escape fractions when using binary stellar populations in their radiation hydrodynamical simulation, while our simulations require lower escape fractions in order to agree with observations. A reason for this disagreement might be that we model the global gas properties of galaxies, while the simulations of \citet{Rosdahl2018} resolve the overall gas density distribution and follow the radiation emitted by the stellar populations in the galaxy.

While the inclusion of binaries may not significantly affect the ionization topology, their inclusion would lead to an increase in certain emission line strengths of galaxies for a constant ionizing escape fraction scenario. We aim to explore this effect in a future work.

\section{Conclusions}
\label{sec_conclusions}

We have developed the semi-analytical/semi-numerical model {\sc astraeus} that self-consistently derives the evolution of galaxies and the reionization of the IGM based on the merger trees and density fields of a DM-only N-body simulation. {\sc astraeus} models gas accretion, star formation, SN feedback, the time and spatial evolution of the ionized regions, accounting for recombinations, \HI fractions and photoionization rates within ionized regions, and radiative feedback. {\sc astraeus} aims at studying the galaxy-reionization interplay in the first billion years.

In this paper, we have focused on the impact of radiative feedback on the interlinked processes of galaxy evolution and reionization. We have considered $6$ radiative feedback and reionization models that cover the physically plausible parameter space. Our models (cf. Table \ref{tab_best_fit_values}) include scenarios (1) where the IGM is heated between $2\times10^4$~K and $4\times10^4$~K, (2) where the gas fraction a galaxy can maintain upon ionization is described by the filtering mass according to \citet{Gnedin2000, Naoz2013} or the Jeans mass, and (3) where the ionizing escape fraction ranges from being constant to scaling with the fraction of gas ejected from the galaxy (i.e. increasing with decreasing halo mass). Comparing the results from these models, we find:
\begin{enumerate}
 \item During the Epoch of Reionization, radiative feedback affects only galaxies with halo masses less than $M_h\sim10^{9.5-10}\Msun$ corresponding to stellar masses less than $M_\star\sim10^{7.5-8}\Msun$ (Fig. \ref{fig_SMFs} and Section \ref{subsec_SMF}) and UV luminosities lower than $M_\mathrm{UV}\gtrsim-16$ (Fig. \ref{fig_UV_LFs} and Section \ref{subsec_UV_LF}). Hence, increasing the strength of radiative feedback, both, the faint end slope of the UV LFs as well as the low-mass end of the SMFs, flatten increasingly as a larger fraction of the IGM becomes ionized
 \item Radiative feedback causes lower gas fractions and suppressed star formation in low-mass galaxies. The strength of radiative feedback increases both with decreasing halo mass and the longer the IGM gas around the galaxy has been ionized: while the temperature of the gas in the galaxy is instantaneously increased when the region becomes ionized, its density adapts to the raised temperature only on the dynamical time scale of the galaxy. The gradual decrease of gas density results in an increasing Jeans mass, steadily lowering the amount of gas the galaxy can maintain. Moderate radiative feedback \citep[e.g.][]{sobacchi2013a} leads to a full suppression of star formation in galaxies with $M_h=10^{8-9}\Msun$, while stronger radiative feedback can reduce the gas content even in galaxies with $M_h=10^{9-10}\Msun$ (Fig. \ref{fig_SFR_radFb} and Section \ref{sec_SFR}).
 \item The radiative feedback strength and hence the degree of star formation suppression depends on the time when the environment of a galaxy becomes ionized; the earlier the IGM around a galaxy is reionized, the more star formation is affected by radiative feedback (Fig. \ref{fig_SFR_radFb} and Section \ref{sec_SFR}). Hence, the patchiness of reionization is imprinted in the star formation histories of low-mass galaxies. However, due to the delayed onset of radiative feedback (Fig. \ref{fig_filtering_masses}) and the rapidness of reionization (Fig. \ref{fig_hist_ion}), for the {\it Photoionization} and {\it Heating} radiative feedback models the differences in observables such as the UV LFs are minor. This is discussed in more detail in an accompanying paper (Ucci et al. 2020).
 \item Radiative feedback does not affect the ionization topology on scales larger than $1-2$~comoving Mpc, as it affects only the star formation in low-mass galaxies during reionization. Consequently, we find the power spectrum of the 21cm signal to be hardly altered on such scales for even the strongest radiative feedback models (Fig. \ref{fig_ps21cm_redshift} and Section \ref{sec_21cm_ps}). Galactic properties such as the ionizing escape fraction $f_\mathrm{esc}$, however, are far more crucial in determining the ionization topology (Fig. \ref{fig_XHI_fields}).
 \item Changing the stellar population synthesis model, i.e. going from stellar populations modelling single stars to those including binaries, has no visible effect on the ionization topology after the escape fraction is tuned to yield reionization histories in agreement with observations.
 \item Our results on the impact of radiative feedback on galactic properties obtained with {\sc astraeus} are in agreement with the findings in radiation hydrodynamical simulations \citep[e.g.][]{ocvirk2016, Wu2019}, such as the galaxies affected by radiative feedback, the degree of star formation suppression in low-mass halos as well as its dependence on the time when a galaxy's environment was reionized (c.f. Section \ref{sec_SFR} and Appendix \ref{app_comparison}). In particular, the results for the {\it Photoionization} or the {\it Early Heating} ($T_0=4\times10^4$~K, $M_c=M_F$) model agree best with the results of radiation hydrodynamical simulations. However, even for those models, we find {\sc astraeus} to have lower baryon fractions in low-mass galaxies compared to radiation hydrodynamical simulations, since {\sc astraeus} accounts only for the kinetic but not temperature effects of SN feedback and not for both as numerical simulations do.
\end{enumerate}

In the following we list some caveats.
Firstly, due to the mass resolution of the underlying N-body simulation, the star formation histories of galaxies with $M_h<10^{8.6}\msun$ are not fully converged. However, we have run our {\sc astraeus} model using an N-body simulation with a $\sim20\times$ better mass resolution, which will allow us to analyse the converged properties of these low-mass galaxies. Although the initial starburst of galaxies at the mass resolution limit could result in a change of the reionization history and topology, this effect is probably negligible, since the aforementioned increased SFR lasts only for a single time step or few really short ($<3$~Myr) time steps and hence the number of additional ionizing photons remains limited.
Secondly, our framework assumes the same low metallicity for all stellar populations and does not account for the evolution of the metallicity in gas and stars, the evolution of the dust content in galaxies, and the associated changes in the stellar populations. This assumption probably begins to break down towards lower redshifts $z\lesssim6$, and we will present a self-consistent treatment of the metallicity evolution within the {\sc astraeus} framework in a follow-up paper (Ucci et al., in prep.).
Thirdly, the redshift of reionization of each grid cell that determines the onset of radiative feedback depends on the grid resolution. In our simulations, a galaxy needs to ionize at least a volume of $\sim0.05$~cMpc$^{-3}$ before its cell is considered reionized. A finer resolution would lead to more accurate results for fainter sources that would then be able to ionize their smaller grid cell earlier. 

Finally, we note that {\sc astraeus} reproduces the key observations at $z\gtrsim5$ using only three mass- and redshift-free parameters and requires much less computation time than comparable radiation hydrodynamical simulation. The underlying code is publicly available\footnote{\url{https://github.com/annehutter/astraeus}}, and written in a modular fashion that allows to incorporate new physics easily. This and especially its short computing time make {\sc astraeus} well equipped to pursue quick and efficient parameter studies, exploring the impact of varying galaxy properties on the ionization topology and their visibility with current and forthcoming telescopes.


\section*{Acknowledgments} 
The authors thank the anonymous referees for their comments that improved the quality of the paper. All the authors acknowledge support from the European Research Council's starting grant ERC StG-717001 (``DELPHI"). PD acknowledges support from the NWO grant 016.VIDI.189.162 (``ODIN") and the European Commission's and University of Groningen's CO-FUND Rosalind Franklin program. 
GY acknowledges financial support from  MICIU/FEDER under project grant PGC2018-094975-C21. We thank Peter Behroozi for creating and providing the {\sc rockstar} merger trees of the {\sc vsmdpl} and {\sc esmdpl} simulations, and AH thanks Manodeep Sinha for discussions about the benefits of local horizontal merger trees. 
The authors wish to thank V. Springel for allowing us to use the L-Gadget2 code to run the different Multidark simulation boxes, including the {\sc vsmdpl} and {\sc esmdpl} used in this work. The {\sc vsmdpl} and {\sc esmdpl} simulations have been performed at LRZ Munich within the project pr87yi. The CosmoSim database (\url{www.cosmosim.org}) provides access to the simulation and the Rockstar data. The database is a service by the Leibniz Institute for Astrophysics Potsdam (AIP). 
This research made use of \texttt{matplotlib}, a Python library for publication quality graphics \citep{hunter2007}; and the Python library \texttt{numpy} \citep{numpy}.

\section*{Data Availability} 

The source code of the semi-numerical galaxy evolution and reionization model within the {\sc astraeus} framework and the employed analysis scripts are available on GitHub (\url{https://github.com/annehutter/astraeus}). The underlying N-body DM simulation, the {\sc astraeus} simulations and derived data in this research will be shared on reasonable request to the corresponding author.

\bibliographystyle{mnras}
\bibliography{delphi}

\appendix

\section{Local-vertical merger trees}
\label{app_merger_trees}

\begin{figure}
 \centering
 \includegraphics[width=0.49\textwidth]{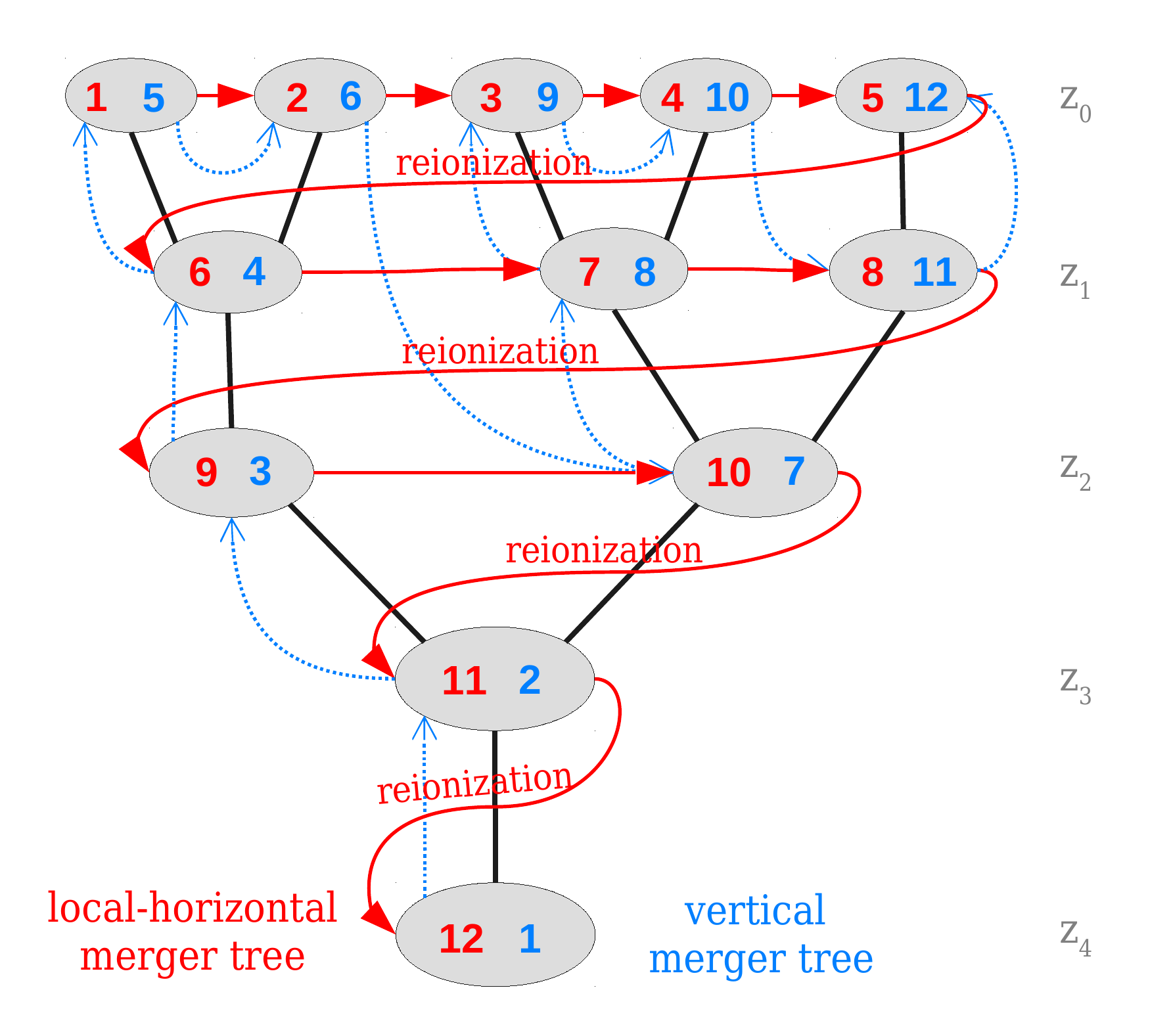}
 \caption{Sketch explaining the structure of local-horizontal merger trees (red) in comparison to traditional vertical merger trees (blue). Local-horizontal merger trees are designed to support both horizontal process, such as reionization, and vertical processes, such as galaxy evolution of a tree.}
 \label{fig_merger_trees}
\end{figure}

Common semi-analytical galaxy evolution models employ vertical merger trees, where the galaxies are ordered on a tree-branch-by-tree-branch basis (see blue numbers and lines in Fig. \ref{fig_merger_trees}). This order allows these model codes to recursively process the evolution of galaxies on a tree-by-tree basis. However, this order does not allow to account for horizontal processes such as reionization (that require redshift-by-redshift processing) without re-running the galaxy evolution model at each redshift step. In order to run the semi-analytical galaxy evolution model only once, we re-order the merger trees to a ``local horizontal" order. While a fully horizontal order is usually referred to as ordering all galaxies by redshift irrespective of their merger tree affiliation, a local horizontal order keeps the galaxies of the same merger tree grouped together, i.e. maintains the tree-by-tree basis, but sorts the galaxies within a merger tree by redshift (see red numbers and lines in Fig. \ref{fig_merger_trees}). This order has the key advantage that it allows us to follow the evolution of single galaxies with a pre-defined reionization redshift  as well as evolving the galaxy population and reionization of the IGM self-consistently and simultaneously.

\section{Impact of the mass resolution of the underlying N-body simulation}
\label{app_resolution}
\begin{figure*}
 \centering
 \includegraphics[width=0.85\textwidth]{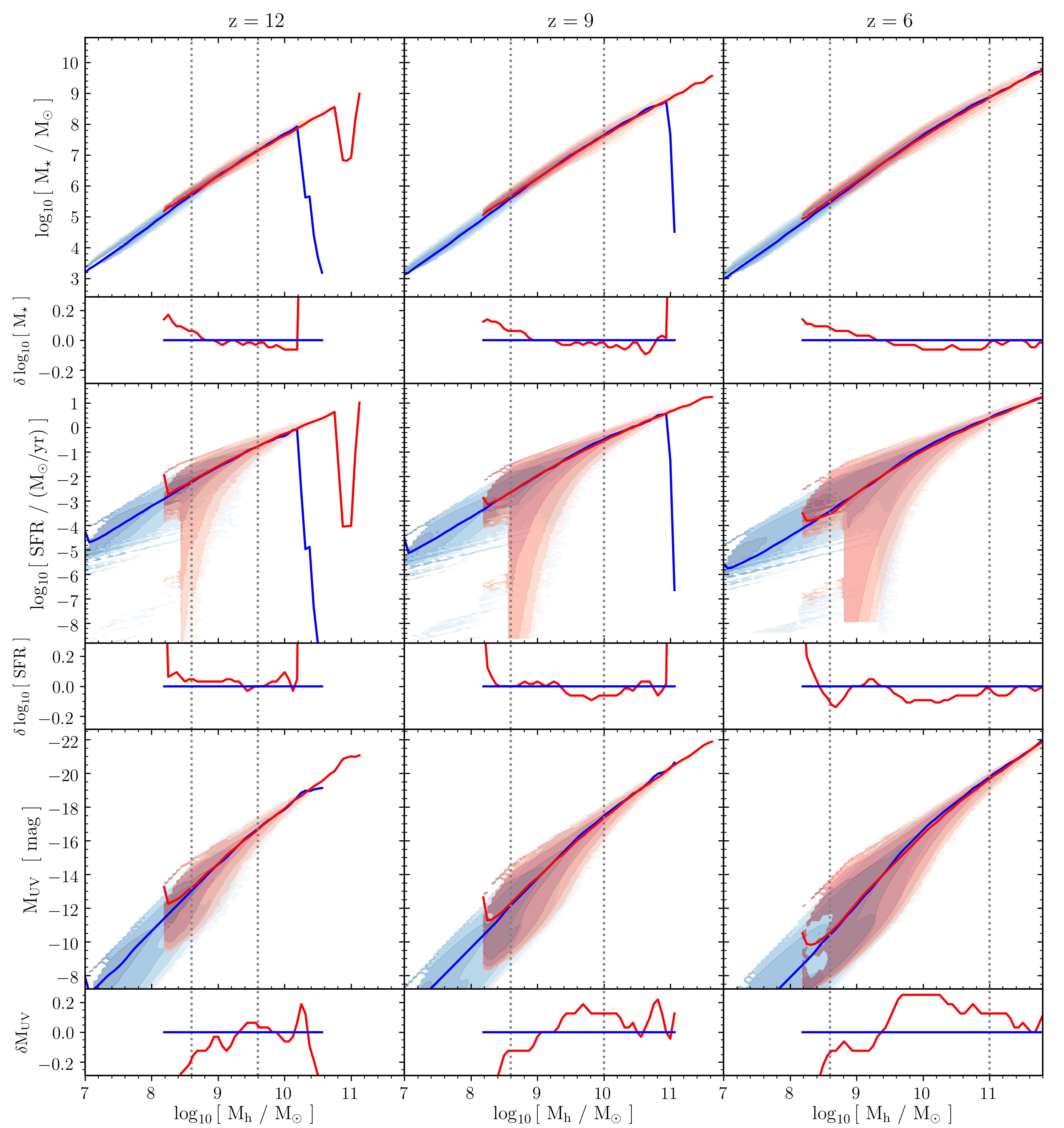}
 \caption{Comparison of the stellar masses (first row), star formation rate (third row) and UV luminosity - halo mass relations (fifth row) from the {\sc vsmdpl} (red) and {\sc esmdpl} (blue) simulation assuming no radiative feedback. Coloured contours indicate the probability density distribution of the respective quantities, while solid coloured lines show the median values in each halo mass bin. The second, fourth, and sixth row show the respective difference to the median relation of the {\sc esmdpl} simulation, whereas the blue line indicated the results for the {\sc esmdpl} and red for the {\sc vsmdpl} simulations. The results based on the two simulations are in agreement between the left vertical grey dotted line (the convergence limit) and the right grey dotted (the cosmic variance limit).} 
 \label{fig_ESMDPL_VSMDPL_comparison_noRadFb}
\end{figure*}

\begin{figure*}
 \centering
 \includegraphics[width=0.85\textwidth]{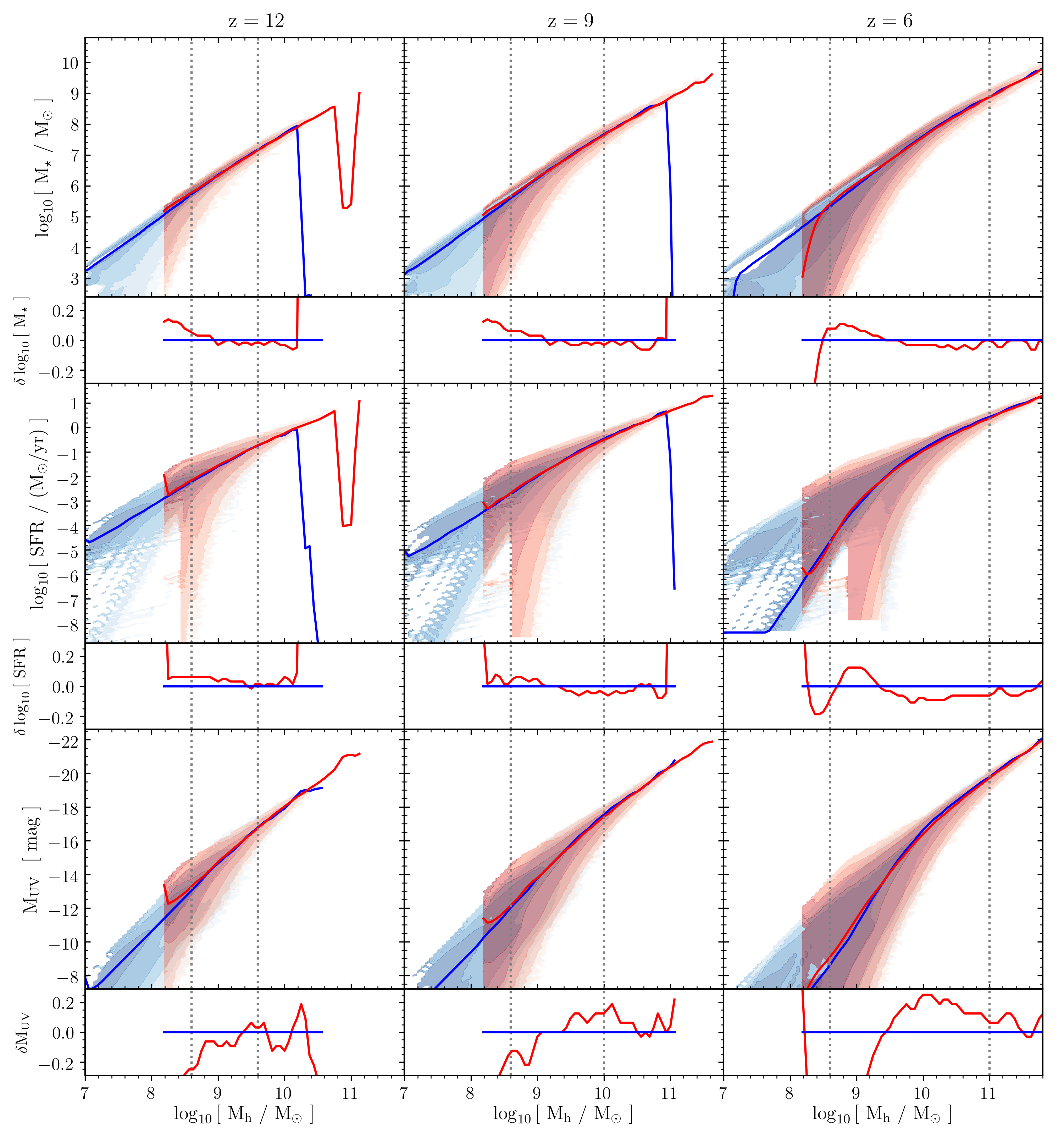}
 \caption{ Comparison of the stellar masses (first row), star formation rate (third row) and UV luminosity - halo mass relations (fifth row) from the {\sc vsmdpl} (red) and {\sc esmdpl} (blue) simulation for the {\it Strong Heating} radiative feedback model. Coloured contours indicate the probability density distribution of the respective quantities, while solid coloured lines show the median values in each halo mass bin. The second, fourth, and sixth row show the respective difference to the median relation of the {\sc esmdpl} simulation, whereas the blue line indicated the results for the {\sc esmdpl} and red for the {\sc vsmdpl} simulations. The results based on the two simulations are in agreement between the left vertical grey dotted line (the convergence limit) and the right grey dotted (the cosmic variance limit).} 
 \label{fig_ESMDPL_VSMDPL_comparison_StrongHeating}
\end{figure*}

In this Appendix we present a resolution study of our {\sc astraeus} model based on the results from the {\sc vsmdpl} (described in Section \ref{subsec_nbody}) and the much more resolved {\it Extremely Small MultiDark Planck} ({\sc esmdpl}) simulation described below. We demonstrate that the {\sc astraeus} model converges for halos with more than $50$ DM particles.

\subsection{{\sc esmdpl} simulation}

The {Extremely Small MultiDark Planck} ({\sc esmdpl}) simulation, with a box size of $64h^{-1}$~cMpc, has been run with $4096^3$ particles and the same cosmological parameters as the {\sc vsmdpl} simulation, $[\Omega_\Lambda, \Omega_m, \Omega_b, h, n_s, \sigma_8]=[0.69, 0.31, 0.048, 0.68, 0.96, 0.83]$. Each DM particle in the simulation has a mass of $3.3\times10^5h^{-1}\msun$, leading to a $20\times$ better mass resolution than in the {\sc vsmdpl} simulation. Halos and subhalos with a minimum of $20$ particles have been identified with the phase-space halo finder {\sc rockstar} \citep{behroozi2013_rs} in each snapshot. From these {\sc rockstar} halo catalogues merger trees have been built using the {\sc consistent trees} method \citet{behroozi2013_trees}. The resulting vertical merger trees have been resorted with {\sc cutnresort} within the {\sc astraeus} pipeline to a local horizontal layout as explained in Appendix \ref{app_merger_trees}.

Due to cosmic variance the number density of the most massive halos in the {\sc esmdpl} simulation drops relative to the corresponding number density in the {\sc vsmdpl} simulation. The lowest halo mass where the halo mass functions of both simulations start to deviate are marked by the right grey dotted lines in Fig. \ref{fig_ESMDPL_VSMDPL_comparison_noRadFb} and \ref{fig_ESMDPL_VSMDPL_comparison_StrongHeating}. 

\subsection{Comparing galactic properties}

We have run our {\sc astraeus} model on the merger trees and density fields of the {\sc esmdpl} and {\sc vsmdpl} simulations for two scenarios: one without accounting for radiative feedback and one where we assume the {\it Strong Heating} model.
In Fig. \ref{fig_ESMDPL_VSMDPL_comparison_noRadFb} and Fig. \ref{fig_ESMDPL_VSMDPL_comparison_StrongHeating} we show the distributions of key galactic quantities, such as the stellar mass, the star formation rate (SFR) and UV luminosity, as a function of halo mass throughout reionization at $z=12$ (left), $9$ (centre) and $6$ (right) for the {\it no radiative feedback} case and the {\it Strong Heating} model, respectively.  
Overall we can see that the distributions and their medians (as a function of the halo mass $M_h$) of the {\sc esmdpl} (blue) and {\sc vsmdpl} (red) simulations are in good agreement. However, close to the mass resolution limit of the {\sc vsmdpl} around $M_h\simeq10^{8.2-8.4}\msun$ we find the properties in the {\sc vsmdpl} simulation to deviate from those in the {\sc esmdpl} simulation, which we briefly discuss in the following.

Newly formed halos have an initial gas mass of $(\Omega_b/\Omega_m M_h)$ and experience an initial starburst in their first time step. This enhanced SFR can indeed be seen at $M_h\simeq10^{8.2}\msun$ for the SFRs in the {\sc vsmdpl} simulation when compared to those in the {\sc esmdpl} simulation (third rows in Fig. \ref{fig_ESMDPL_VSMDPL_comparison_noRadFb} and \ref{fig_ESMDPL_VSMDPL_comparison_StrongHeating}). However, upon this initial starburst, SN feedback will eject all of the gas in these SN feedback-limited galaxies in the second time step (and very latest in the third time step at $z\gtrsim20$ when time steps are close to $3$~Myrs). The gas available for star formation in subsequent time steps is then given by the amount of gas that these galaxies accrete. Hence, the SFRs converge within a couple of time steps, which is why the median SFR in the {\sc vsmdpl} simulation approaches the median SFR in the {\sc esmdpl} simulation already around $10^{8.4}\msun$ across all redshifts.

The initial starburst also causes the median stellar mass in the {\sc vsmdpl} simulation to be slightly higher than in the {\sc esmdpl} simulation at the mass resolution limit of the {\sc vsmdpl} simulation at $M_h\simeq10^{8.2}\msun$ (see first rows in Fig. \ref{fig_ESMDPL_VSMDPL_comparison_noRadFb} and \ref{fig_ESMDPL_VSMDPL_comparison_StrongHeating}). However, since the SFRs converge within a couple of time steps, we also find the median stellar mass to converge around $M_h\simeq10^{8.5}\msun$. This trend occurs because the total gas and DM masses that are accreted by these low-mass galaxies exceed the gas and halo masses at the time of their formation. Hence, the stellar mass formed in the initial starburst becomes smaller than all the stellar mass formed after (when the SFR has already converged).

The UV luminosity (fifth rows in Fig. \ref{fig_ESMDPL_VSMDPL_comparison_noRadFb} and \ref{fig_ESMDPL_VSMDPL_comparison_StrongHeating}) of a galaxy traces predominantly its current SFR but is also sensitive to its SFH (and hence to its stellar mass). For this reason the increased SFR at the mass resolution limit ($M_h\simeq10^{8.2}\msun$) of the {\sc vsmdpl} simulation translates into a higher UV luminosity for these low-mass galaxies compared to the {\sc esmdpl} simulation. However, since the UV luminosity of a galaxy also depends on previous recent star formation, we find the median UV luminosities of the galaxies in the {\sc vsmdpl} simulation to converge at slightly higher halo masses than the medians of their SFRs and stellar masses, around $M_h\simeq10^{8.6}\msun$ (see Fig. \ref{fig_ESMDPL_VSMDPL_comparison_noRadFb}). We note that the seemingly larger deviations in the UV luminosity are still smaller than the observational uncertainties at these redshifts.

Our comparison of galactic properties derived from the {\sc vsmdpl} and {\sc esmdpl} simulations suggests that the {\sc astraeus} model converges in the {\sc vsmdpl} simulation for halos with masses of $M_h\geq10^{8.6}\msun$, corresponding to halos with $N_p\geq50$ DM particles. The reason that our model converges rather quickly is its direct linking of star formation processes to the underlying gravitational potential which is traced by the halo mass.

\section{Comparison with other simulations}
\label{app_comparison}

We compare our simulation results with those from other semi-analytic galaxy evolution models that self-consistently include reionization \citep{Mutch2016, Seiler2019} and radiation hydrodynamic simulations \citep{Hasegawa2013, Gnedin2014, Pawlik2015, Wu2019, Ocvirk2018, Katz2019}.

\subsection{Semi-analytical/semi-numerical galaxy and reionization models}
\label{subapp_comparison_sams}

Comparing our results to those of semi-analytic galaxy evolution models, we find that the overprediction of low stellar mass sources with $M_\star\lesssim10^9\Msun$ ($10^8\Msun$) at $z\simeq5$ ($6$) can also be found in {\sc meraxes} \citep{Mutch2016}, and similarly at $z\simeq6$ in {\sc rsage} \citep{Seiler2019}. Furthermore, the global star formation rate density at $z\simeq4-7$ produced by {\sc meraxes} \citep{Qiu2019} are in agreement with our findings and the dust-corrected result of \citet{Bouwens2015}. Since {\sc meraxes} also models radiative feedback by the filtering mass fitting function given in \citet{sobacchi2013a} and the Universe reionizes at similar redshifts, the maximum gas fraction of a halo $f_g\Omega_b/\Omega_m$ agrees well with our findings \citep[see Fig. 1 in][]{Qin2019}, and so do the stellar-to-halo mass distributions \citep[c.f. fiducial model in][]{Mutch2016}. 

\subsection{Radiation hydrodynamical simulations}
\label{subapp_comparison_simulations}

Comparing our results, particularly star formation rates and gas fractions, with radiation hydrodynamical simulations, we find them to agree in their overall trends but to differ in their amplitudes. In the following, we discuss possible reasons for the similarities and differences:

\subsubsection{Total star formation rate density (SFRD)}
\label{subsubapp_comparison_SFRD}
We find the total star formation rate density in radiation hydrodynamical simulations to be lower than that of our semi-numerical simulations at $z\gtrsim8-9$ \citep[c.f.][]{Pawlik2015, ocvirk2016, Ocvirk2018, Hasegawa2013, Wu2019}. Although these simulations include the same physical processes (gas accretion, star formation, SN feedback), they seem to produce less low-mass halos or low-mass halos with lower star formation rates, particularly given that the star formation rate per halo as a function of the halo mass agrees reasonably well with our findings \citep[see also Fig. A1 in][]{ocvirk2016}. There may be multiple reasons for this: 

Firstly, due to hydrostatic pressure preventing gas from collapsing, the collapse or growth of dark matter halos of a mass $M_h$ is delayed or suppressed in hydrodynamical simulations compared to collisionless N-body DM-only simulations \citep[e.g.][]{Sawala2013, Velliscig2014, Schaller2015, Khandai2015, Bocquet2016, Qin2019}. This effect is stronger at higher redshifts, as the higher mean gas density `weakens' the gravitational potentials of collapsing halos\footnote{We note that including SN and reionization feedback further suppresses the growth of dark matter halos, because these feedback processes remove gas or prevent further accretion. This effect becomes stronger towards lower mass halos with weaker gravitational potential.}. Indeed, \citet{Qin2019} find the masses of dark matter halos in their hydrodynamical simulations to be about a factor $2$ at $z\simeq15$ and $\sim 1.3$ at $z\simeq5$ lower than in DM-only N-body simulations. Hence, particularly at high redshifts, the number of DM halos and hence galaxies will be lower in radiation hydrodynamical simulations than in our semi-analytical model based on a DM-only N-body simulation. 

Secondly, \citet{Qin2019} point out that the baryon fraction of collapsed halos is always less than $\sim90\%$ of the cosmological ratio $\Omega_b/\Omega_m$ and does not evolve with redshift when radiative feedback from reionization is not included. Hence, there is less gas available for star formation. Both effects, the decrease in number of dark matter halos and the reduced baryon fraction in halos cause an increasingly lower SFRD towards higher redshifts.

Thirdly, the delayed collapse or suppressed growth of dark matter halos becomes stronger for halo masses closer to the resolution limit. \citet{Qin2019} find the baryon to dark matter ratio to saturate for halos with $\gtrsim10^3$ particles. Hence, the higher the dark matter mass of the radiation hydrodynamical simulation, the stronger are low-mass halos affected. Indeed, we find that the SFRD increasingly decreases towards higher redshifts, as the DM particle mass in the simulation increases, from \citet{Hasegawa2013} with $m_\mathrm{DM}=2.5\times10^5\Msun$ via \citet{Ocvirk2018} with $m_\mathrm{DM}=4.1\times10^5\Msun$ to \citet{Pawlik2015} with $m_\mathrm{DM}=10^7\Msun$. However, the adopted star formation and stellar feedback descriptions can dilute this trend: e.g. in contrast to the other works, \citet{Hasegawa2013} assume a Salpeter instead of a Chabrier IMF, reducing the number of SN formed per stellar mass and boosting star formation\footnote{We note also the implementation of SN feedback affects results. As has been shown by \citet{DallaVecchia2012} a distribution of the SN energy to neighbouring particles results in distributing the energy to too much mass, leading to a smaller increase in temperature, a shorter cooling time, and a weaker stellar feedback causing increased star formation (overcooling problem). \citet{Hasegawa2013} use this description, which possibly contributes to their higher SFRD.}. \citet{Pawlik2015} do not resolve the low-mass sources and needs to boost their star formation efficiency to produce the sufficient stellar mass. \citet{Wu2019} show - despite a DM particle mass of $m_\mathrm{DM}=1.4\times10^6\Msun$ - the lowest SFRD at higher redshifts; in contrast to the other mentioned works, their density threshold for star formation is given by an absolute physical density and not a relative one to the mean, increasingly reducing star formation towards higher redshifts.

\subsubsection{Star formation rate per halo}
\label{subsubapp_comparison_SFR}
Comparing the SFR as a function of halo mass, we find our results to be in agreement with the results from radiation hydrodynamical simulations presented in \citet{Pawlik2015}, \citet{Wu2019} and \citet{Ocvirk2018}; \citet{Hasegawa2013} show SFRs that are about 0.5-1 dex higher, possibly due to their SN feedback implementation and an IMF that results in a lower SN rate per unit stellar mass formed. In particular, after most of the Universe has been ionized, all simulations find a drop in the SFR below $~10^{9-9.5}\Msun$, which is in agreement with our {\it Photoionization} model. \citet{Ocvirk2018} find a slightly stronger suppression in low mass halos, corresponding to a radiative feedback model that lies between our {\it Photoionization} and {\it Strong Heating} models. 

In \citet{Wu2019}, we see the same effect that we find in our SFR histories: switching the UVB on at $z\sim10.7$ causes an early ($z>6$) suppression of star formation in low mass halos. However, when deriving the radiation field self-consistently, the suppression becomes only visible towards lower redshifts. As the Universe becomes more ionized, more galaxies are subject to radiative feedback, and the more time has been passed since a galaxy's environment has been ionized and adapted its gas density to the new temperature.
Our SFR histories also echo the results in \citet{Dawoodbhoy2018}, which are based on the CoDaI simulation \citep{ocvirk2016}. In CodaI, the Universe reionizes just at $z\simeq4.6$ and consequently lower SFRs for halos with $M_h<10^9\Msun$ are found compared to CoDaII \citep{Ocvirk2018} and our simulations. However, they find the same trends of the SFR with halo mass and reionization redshift: SFR in low mass halos ($M_h=10^{8-9}\Msun$) is suppressed immediately upon reionization, SFR in mid mass halos ($M_h=10^{9-10}\Msun$) increases upon reionization but drops later in time, and SFR in massive halos ($M_h>10^{10}\Msun$) is not affected by radiative or SN feedback and continues to increase with time. The mass ranges for these three regimes shift to higher masses as the strength of the radiative feedback is increased, allowing us to pinpoint their radiative feedback strength to lie between our {\it Photoionization} and {\it Strong Heating} models.

\begin{figure*}
 \centering
 \includegraphics[width=0.95\textwidth]{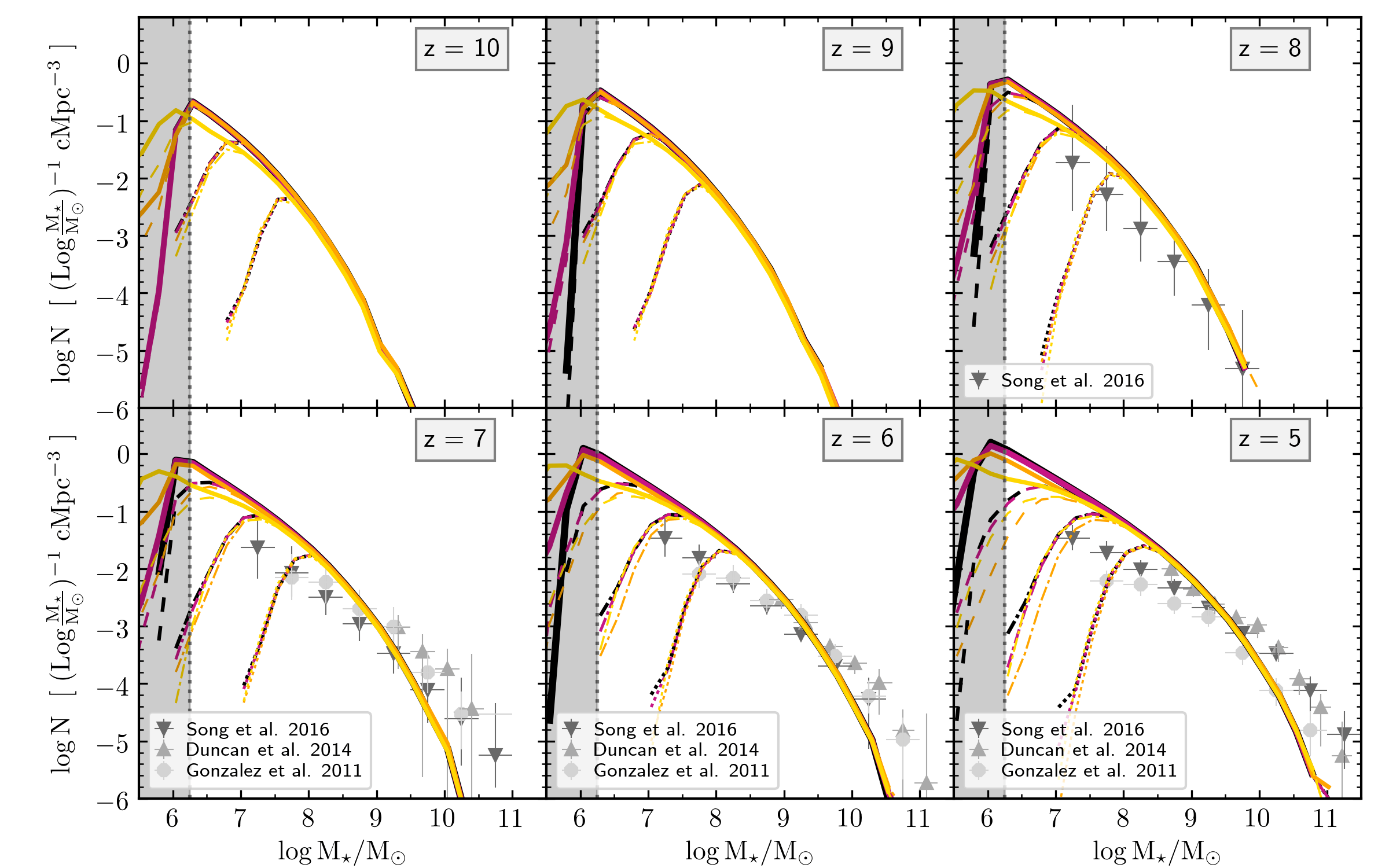}
 \caption{Stellar mass functions (SMFs) at $z \sim 5-10$ using the best-fit parameters noted in Table \ref{tab_best_fit_values} and accounting for all galaxies in halos with $M_h\geq10^{8.6}\msun$. In each panel, the different line types show SMFs using different magnitude limits: all galaxies (solid lines) and galaxies brighter than $\mathrm{M_{UV}}\leq-13$ (dashed lines), $\mathrm{M_{UV}}\leq-15$ (dot-dashed lines) and $\mathrm{M_{UV}}\leq-17$ (dotted lines). In each panel, we show results for the different radiative feedback models studied in this work: {\it Minimum} (black lines), {\it Weak Heating} (blue lines), {\it Photoionization} (violet lines), {\it Early Heating} (red lines), {\it Strong Heating} (orange lines) and {\it Jeans Mass} (yellow lines). The grey shaded area marks the halo masses that might be affected by the resolution limit of the underlying N-body simulation. Finally, the grey points indicate the observational data collected, as marked in the panels for $z\sim 5-8$.  Grey points indicate the observational data points from \citet{Song2016}, \citet{Duncan2014} and \citet{Gonzalez2010} as marked.}
 \label{fig_SMFs}
\end{figure*}

\subsubsection{Baryon or gas fractions:}
\label{subsubapp_comparison_baryon_fraction}

It is important to note that our stellar feedback model represents an upper limit of SN feedback on gas in galaxies: firstly, our model does not account for the cooling of gas within galaxies, and secondly, the gas energised by SN is fully ejected from the halo. This is different to the SN feedback prescriptions in hydrodynamical simulations where gas heated by SN can cool (thermal SN feedback) or gas accelerated by SN energy does not reach the edge of the halo (kinetic SN feedback). 
Hence, in our model, more gas leaves the galaxy than in hydrodynamical simulations, and consequently we find lower gas or baryon fractions in galaxies that are located in lower mass halos. It is interesting to notice that if we consider the gas fraction $f_g$ in Equation \ref{eq_MgasIni}, which is purely due to radiative feedback and given by the respective filtering mass, we find the {\it Photoionization} or a {\it Early Heating} ($T_0=2\times10^4$~K, $M_c=M_F$) model to agree best with the findings at $z\simeq6-7$ in \citet{Hasegawa2013, Okamoto2008, Gnedin2014}\footnote{We note that \citet{Gnedin2014} show the mean and not median gas fraction.}; \citet{Pawlik2015} and the UVB model in \citet{Wu2019}. This finding indicates that, indeed, SN feedback in the aforementioned hydrodynamical simulations is not strong enough to eject the gas from the halo that has been heated or accelerated by SN explosions (but is able to suppress star formation by either heating the gas or decreasing the gas density in star formation sites through winds).

\subsection{Stellar mass functions}
\label{subsec_SMF}

\begin{figure}
 \centering
 \includegraphics[width=0.49\textwidth]{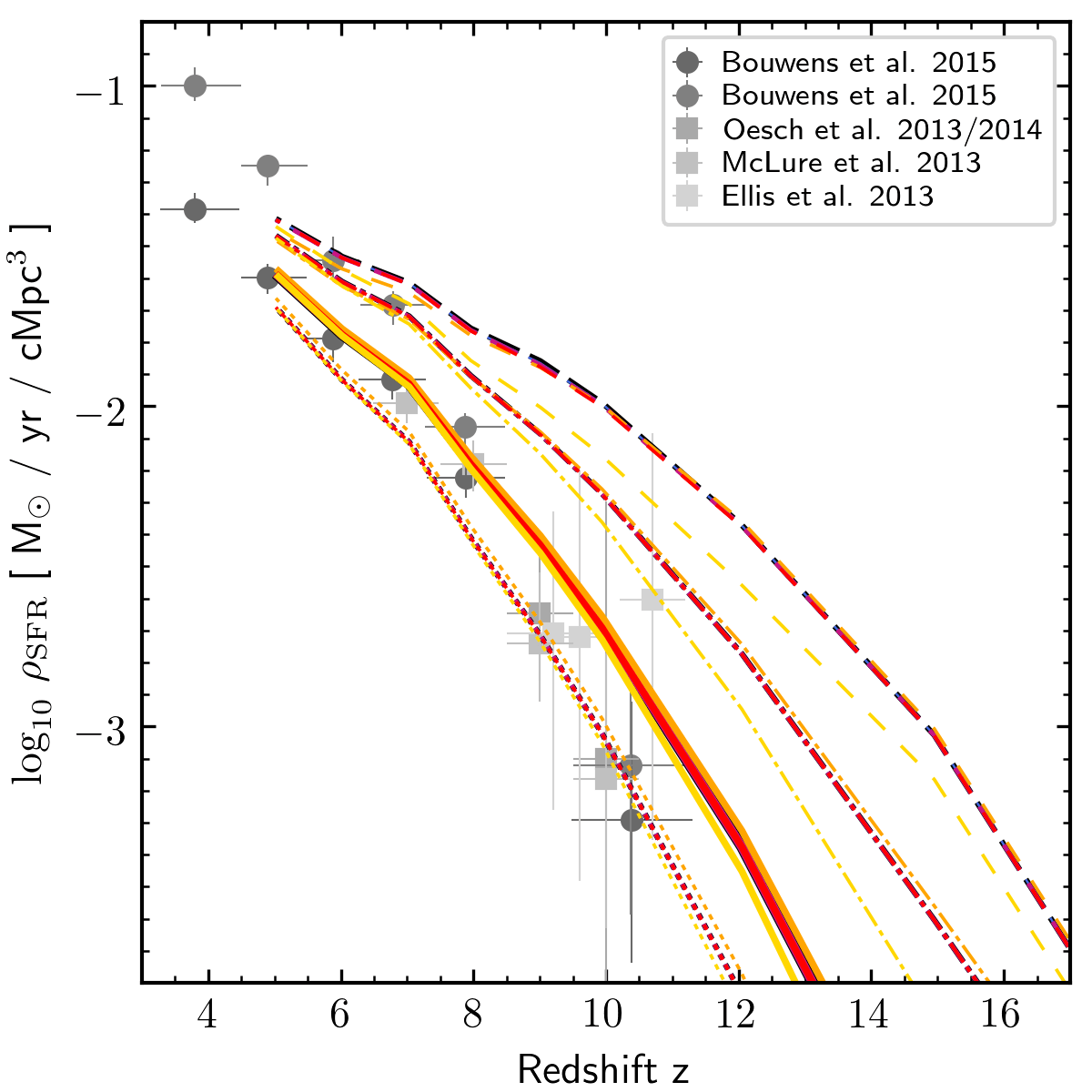}
 \caption{Star formation rate densities (SFRDs) of our best fit models. Dotted, solid, dash-dotted and dashed lines show the SFRDs for all galaxies in the simulation brighter than $\mathrm{M_{UV}}<-18$, $-17$, $-15$ and $-13$, respectively. The different lines correspond to our radiative feedback models: {\it Minimum} (black line), {\it Weak Heating} (blue line), {\it Photoionization} (violet line), {\it Early Heating} (red line), {\it Strong Heating} (orange line) and {\it Jeans Mass} (yellow line). Grey points indicated the observational data collected by \citet{Bouwens2015}, \citet{Oesch2013}, \citet{Oesch2014}, \citet{McLure2013} and \citet{ Ellis2013} for $M_\mathrm{UV}\leq-17$.}
 \label{fig_SFRD}
\end{figure}

\begin{figure}
 \centering
 \includegraphics[width=0.49\textwidth]{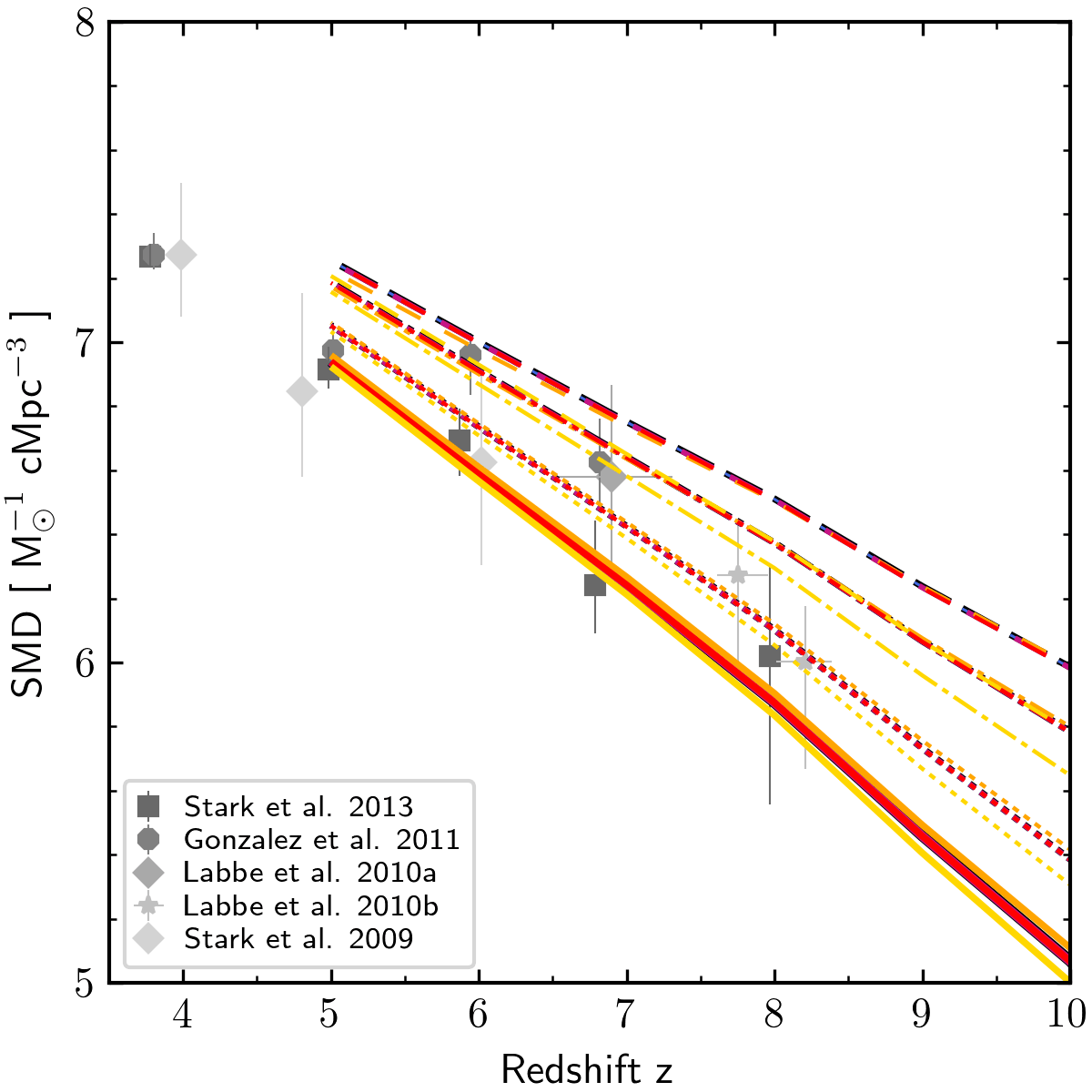}
 \caption{Stellar mass densities (SMDs) of our best fit models. Solid, dotted, dash-dotted and dashed lines show the SMDs for all galaxies in the simulation brighter than $\mathrm{M_{UV}}<-18$, $-17$, $-15$ and $-13$, respectively. The different lines correspond to our radiative feedback models: {\it Minimum} (black line), {\it Weak Heating} (blue line), {\it Photoionization} (violet line), {\it Early Heating} (red line), {\it Strong Heating} (orange line) and {\it Jeans Mass} (yellow line). Grey points indicate the observational data points from \citet{Gonzalez2011}, \citet{Labbe2010a}, \citet{Labbe2010b}, \citet{Stark2009} and \citet{Stark2013} for $M_\mathrm{UV}\leq-18$.}
 \label{fig_SMD}
\end{figure}

We now discuss the stellar mass functions (SMFs) at $z=5-10$ obtained for the different feedback models used in this work, as shown in Fig. \ref{fig_SMFs}. We find that the overall normalisation of the SMF increases with decreasing redshift, as new galaxies form and existing ones grow. As expected, the SMF at the massive end ($M_h\gtrsim10^{9.5}\msun$ or $M_\star\gtrsim10^{7}\msun$) increases with decreasing redshift as these galaxies assemble mass through star formation (at the maximum threshold efficiency $f_*$) and mergers. The low-mass end ($M_\star\lesssim10^8\msun$) flattens as redshift decreases: this is due to a combination of SN and radiative feedback driven decrease in the star formation efficiency, as well as low-mass galaxies moving into higher mass bins with time.
When accounting only for supernova feedback (i.e. the {\it Minimum model} at $z\gtrsim5.8$), the change in the low-mass slope is moderate. Analogous to the UV LF, adding radiative feedback enhances the flattening of the low-mass slope with decreasing redshift as a larger number of low-mass halos are feedback suppressed. The flattening increases as radiative feedback becomes stronger (i.e. the characteristic mass $M_c$ increases), going from the {\it Photoionization} to the {\it Strong Heating} model), i.e. as low-mass galaxies become increasingly gas-poor and less efficient in forming stars. 

Further, similar to the UV LF, the maximum stellar mass $M_{\star,s}$ suppressed by radiative feedback, i.e. the mass at which the low-mass slope deviates from that of the {\it Minimum} model, increases with the radiative feedback strength. For example, it is $\sim10^{8}\Msun$ for the {\it Strong Heating} model and would be $M_{\star,s}\simeq10^{6}\Msun$ for the {\it Photoionization} model (inferred from the corresponding SMFs derived from the {\sc esmdpl} simulation discussed in Appendix \ref{app_resolution}) at $z\lesssim7$, respectively. Analogous to $M_\mathrm{UV,s}$, $M_{\star,s}$ traces the highest possible radiative feedback characteristic mass at a given redshift, which naturally increases with increasing radiative feedback strength. Again the {\it Jeans Mass} model constitutes an exception: here $M_{\star,s}$ hardly evolves with cosmic time and corresponds consistently to the stellar mass in a halo of Jeans mass $M_J(z)$. This difference becomes particularly obvious at $z=5$ when the radiative feedback characteristic mass $M_c$ of the {\it Strong Heating} model exceeds that of the {\it Jeans Mass} model. This results in the SMFs (for $M_\mathrm{UV}\leq -13$ and $M_\mathrm{UV}\leq -15$ in Fig. \ref{fig_SMFs}) turning over at higher masses whilst showing a weaker suppression of stellar mass at the low-mass end ($M_\star\lesssim10^7\msun$).

Our model results of the SMFs at $z=5-10$ are in agreement within the uncertainties of the observations. However, from Fig. \ref{fig_SMFs} we can see that our model SMFs tend to overpredict the low-mass and underpredict the high-mass end of the observed SMF, despite our UV LFs at $z=5-10$ being in good agreement with the observations (although slightly overpredicting the bright end as noted in Sec. \ref{subsec_UV_LF}). 
There may be multiple reasons for these trends: firstly, from the modelling side, we do not account for dust attenuation when computing the UV luminosities - this can underestimate the star formation efficiency $f_\star$ and the resulting stellar masses. 
Secondly, the observational data points are subject to modelling uncertainties when deriving the stellar masses from broadband fluxes via spectral energy distribution (SED) fitting: on the one hand, the uncertainty of emission lines contributing to the broadband flux in the spectra of high-redshift galaxies could alter the derived observational SMF \citep{Song2016} but also assumptions on the assumed star formation histories, dust contents and metallicities.

\section{Star formation rate density \& stellar mass density}

For all our models, we show the star formation rate densities (SFRDs) and stellar mass functions for different UV luminosity selection criteria, i.e. $M_\mathrm{UV}<-13$, $-15$, $-17$, $-18$ in Fig. \ref{fig_SFRD} and \ref{fig_SMD}, respectively. Applying the same selection criterion as the observations (i.e. $M_\mathrm{UV}<-17$), we find all our models to be in good agreement with the observational data points for both the SFRDs and SMDs, tracking the overall increase in star formation and stellar mass with cosmic time as new galaxies form and existing ones continuously grow. 
We can see that the SFRDs and SMDs of our models agree well with each other, since firstly the suppression of star formation by radiative feedback is mostly found in low-mass and UV faint galaxies where star formation is already limited by SN feedback, and secondly the reduction of star formation by radiative feedback is time delayed. Only the {\it Jeans Mass} model deviates when faint galaxies are included: radiative feedback suppresses star formation instantaneously when the environment of the galaxy becomes ionized. In addition, it affects only galaxies in low-mass halos, and hence reduces the SFRD and SMDs only for $M_\mathrm{UV}>-13$ and $-15$.

\bsp	
\label{lastpage}
\end{document}